\documentstyle[aps,prb,epsfig,psfig,floats,color]{revtex}

\begin{document}                                                             
\newcommand{\beq}{\begin{equation}}
\newcommand{\eeq}{\end{equation}}
\newcommand{\bea}{\begin{eqnarray}}
\newcommand{\eea}{\end{eqnarray}}
\newcommand{\dfrac}{\displaystyle\frac}                                    
\newcommand{\disp}{\displaystyle}                                    
\newcommand{\mbf}{\mathbf}
\newcommand{\dint}{\displaystyle\int}
\renewcommand{\u}{\underline}                                        
\renewcommand{\o}{\overline}                                        
\newcommand{\Ptwo}{{\mathcal{P}}_2}
\newcommand{\Ptwon}{\mathcal{P}_{2n}}
\newcommand{\bi}{\begin{itemize}}
\newcommand{\ei}{\end{itemize}}
\newcommand{\thint}{\widetilde{d\theta}}
\newcommand{\thprint}{\widetilde{d\theta'}}
\newcommand{\angint}{\int \thint\,}
\newcommand{\totint}{\int dl\,\thint\,}
\newcommand{\lint}{\int dl\,}
\newcommand{\fexc}{\tilde{f}}
\newcommand{\fmom}{f_{\rm{ mom}}}
\newcommand{\muexc}{\tilde\mu}
\newcommand{\fid}{f_{\rm{id}}}
\newcommand{\rhoa}{\rho^{(a)}}
\newcommand{\va}{v^{(a)}}
\newcommand{\rhob}{\rho^{(b)}}
\newcommand{\vb}{v^{(b)}}
\newcommand{\parent}{\rho^{(0)}}
\newcommand{\normparent}{P^{(0)}}
\newcommand{\Ilm}{\mathcal{I}}
\newcommand{\phiparent}{\phi^{(0)}}
\newcommand{\bfig}{\begin{figure}[htb]\begin{center}}
\newcommand{\efig}{\end{center}\end{figure}}
\newcommand{\eqref}[1]{(\ref{#1})}
\newcommand{\partf}{\mathcal{Z}}
\newcommand{\Nalpha}{N^{(\alpha)}}
\newcommand{\Valpha}{V^{(\alpha)}}
\newcommand{\valpha}{v^{(\alpha)}}
\newcommand{\rhoalpha}{\rho^{(\alpha)}}
\newcommand{\ssch}{\sigma_{\rm S}}

\newtheorem{thm}{Theorem}     [section]                                    
\newtheorem{definition}{Def.}   [section]                                    
\newtheorem{obs}{Remark}  [section]
\newcommand{\vectr}{{\text{\boldmath$r$}}}
\newcommand{\vectp}{{\text{\boldmath$p$}}}


\newcommand{\be}{\begin{equation}}
\newcommand{\ee}{\end{equation}}
\newcommand{\bestar}{\[}
\newcommand{\eestar}{\]}
\newcommand{\beastar}{\begin{eqnarray*}}
\newcommand{\eeastar}{\end{eqnarray*}}
\newcommand{\nn}{\nonumber\\}
\newcommand{\lav}{\left\langle}
\newcommand{\rav}{\right\rangle}

\newcommand{\half}{\frac{1}{2}}
\newcommand{\eqm}[2]{~(\ref{#1}-\ref{#1})}
\newcommand{\eq}[1]{~(\ref{#1})}
\newcommand{\eqq}[2]{~(\ref{#1},\ref{#2})}
\newcommand{\eqqq}[3]{~(\ref{#1},\ref{#2},\ref{#3})}
\newcommand{\sgn}{\mbox{sgn}}
\newcommand{\erf}{\mbox{erf}}
\newcommand{\order}{{{\mathcal O}}}

\newcommand{\ie}{{\it i.e.}}
\newcommand{\eg}{{\it e.g.}}
\newcommand{\eqrefm}[2]{~(\ref{#1}-\ref{#1})}


\newcommand{\rhol}{\rho(l)}
\newcommand{\rhotl}{\rho(l,\theta)}
\renewcommand{\pl}{P(l)}
\newcommand{\x}{_{\rm x}}
\newcommand{\y}{_{\rm y}}
\newcommand{\z}{_{\rm z}}

\newcommand{\pdl}{P_l(d)}
\newcommand{\pxl}{P_l({\rm x})}
\newcommand{\pyl}{P_l({\rm y})}
\newcommand{\pzl}{P_l({\rm z})}

\newcommand{\rhox}{\rho\x}
\newcommand{\rhoy}{\rho\y}
\newcommand{\rhoz}{\rho\z}
\newcommand{\mom}{\phi}
\newcommand{\dmom}{\Delta}
\newcommand{\intl}{\int\!\!dl\,}
\newcommand{\sd}{\sum_d}
\newcommand{\laml}{\kappa(l)}
\newcommand{\pa}{^{(\al)}}
\newcommand{\pn}{^{(0)}}
\newcommand{\rhoal}{\rho\pa(l)}
\newcommand{\rhon}{\rho^{(0)}}
\newcommand{\pnl}{P^{(0)}(l)}
\newcommand{\rhonl}{\rho\pn(l)}
\newcommand{\rhondl}{\rho_d\pn(l)}
\newcommand{\al}{a}
\newcommand{\bet}{b}
\newcommand{\lamb}{\alpha}
\newcommand{\apar}{\lamb_\parallel}
\newcommand{\aper}{\lamb_\perp}
\newcommand{\K}{K}
\newcommand{\I}{^{\rm I}}
\newcommand{\N}{^{\rm N}}
\newcommand{\m}{m}
\newcommand{\rl}{r(l)}
\newcommand{\ptl}{P(\theta|l)}
\newcommand{\rtl}{\rho(l,\theta)}
\newcommand{\ptlpr}{P(\theta'|l')}
\newcommand{\psit}{\psi(\theta)}
\newcommand{\psitpr}{\psi(\theta')}
\newcommand{\iptl}{P\I(\theta|l)}
\newcommand{\nptl}{P\N(\theta|l)}
\newcommand{\gt}{g(\theta)}
\newcommand{\gnot}{g(0)}
\renewcommand{\l}{(l)}
\renewcommand{\t}{(\theta)}
\newcommand{\tpr}{(\theta')}
\newcommand{\tl}{(\theta,l)}
\newcommand{\llangle}{\left\langle}
\newcommand{\rrangle}{\right\rangle}
\newcommand{\Ks}{\widetilde K}
\newcommand{\hs}{\widetilde h}
\newcommand{\Ps}{\widetilde P}
\newcommand{\rhoeff}{\rho_{\rm eff}}

\twocolumn[\hsize\textwidth\columnwidth\hsize\csname @twocolumnfalse\endcsname

\title{Isotropic-nematic phase equilibria in the Onsager theory of
  hard rods with length polydispersity}
\author{Alessandro Speranza, Peter Sollich}

\address{Department of Mathematics, King's College London,
Strand, London WC2R 2LS, U.K.
Email: peter.sollich@kcl.ac.uk}

\maketitle

\begin{abstract}
We analyse the effect of a continuous spread of particle lengths on
the phase behavior of rodlike particles, using the Onsager theory of
hard rods. Our aim is to establish whether ``unusual'' effects such as
isotropic-nematic-nematic (I-N-N) phase separation can occur even for
length distributions with a single peak. We focus on the onset of I-N
coexistence. For a log-normal distribution we find that a finite upper
cutoff on rod lengths is required to make this problem well-posed. The
cloud curve, which tracks the density at the onset of I-N coexistence
as a function of the width of the length distribution, exhibits a
kink; this demonstrates that the phase diagram must contain a
three-phase I-N-N region.
%
%
Theoretical analysis shows that in the limit of large cutoff the cloud
point density actually converges to zero, so that phase separation
results at any nonzero density; this conclusion applies to all length
distributions with fatter-than-exponentail tails. Finally we consider
the case of a Schulz distribution, with its exponential
tail. Surprisingly, even here the long rods (and hence the cutoff) can
dominate the phase behaviour, and a kink in the cloud curve and I-N-N
coexistence again result. Theory establishes that there is a nonzero
threshold for the width of the length distribution above which these long rod
effects occur, and shows that the cloud and shadow curves approach
nonzero limits for large cutoff, both in good agreement with the
numerical results.
\end{abstract}

\vspace*{0.5cm}
]

\section{Introduction}

Rodlike particles such as Tobacco Mosaic Virus (TMV) in dilute
suspension are known to exhibit a phase transition with increasing
density between an isotropic phase (I) with no orientational or
translational order and a nematic phase (N) where the rods point on
average along a preferred
direction~\cite{BawPir37,BawPir38,BawPirBerFan36,BerFan37,BerFan41}.
The main theoretical approach formulated to predict this phenomenon is
the Onsager theory of hard rods~\cite{Onsager49}. Onsager assumed that
the only interaction between the solute particles is of hard
core type. The particles are modeled as perfectly rigid long, thin
rods; non-rigidity as well as possible long-range attractive
potentials are neglected. Crucially, Onsager showed that the virial
expansion truncated after the first non-trivial contribution becomes exact
in the limit of long thin rods (the ``Onsager limit''), \ie, for
$D/L_0\rightarrow 0$ where $D$ is the diameter and $L_0$ the length of
the rods. The free energy then assumes a very simple form, because the
second virial coefficient is just the excluded volume of two
rods. The Onsager limit does however constrain the
theory to low densities of order $\rho\sim\order(1/DL_0^2)$, and
phases such as smectics which occur at higher density cannot be
predicted.

In order to express the distribution of the non-conserved rod
orientations, Onsager introduced the probability $P(\Omega)$ of
finding a rod pointing along the direction $\Omega$. Minimization of
the free energy with respect to $P(\Omega)$ results in a
self-consistency equation for $P(\Omega)$. Solving this in principle
reduces the free energy to a function of the density $\rho$ only, so
that phase coexistences can be found by a standard double tangent
construction. In order to avoid the complexity of the numerical
solution~\cite{HerBerWin84,LeeMey86} of the self-consistency equation,
Onsager used a simple one-parameter variational trial form for
$P(\Omega)$. Using this method Onsager~\cite{Onsager49}
and, two years later Isihara~\cite{Isihara51}, were able to estimate
the density at which the I-N phase transition occurs for different
particle shapes. A similar approach was used by Odijk~\cite{Odijk86},
with a Gaussian trial function for $P(\Omega)$. However, the
numerically exact solution had by then already been obtained by Kayser
\& Ravech\'e~\cite{KayRav78}. An alternative method, based on an
expansion of the angular part of the excluded volume in terms of
Legendre polynomials~\cite{KayRav78}, was used by Lekkerkerker {\em et
al.}~\cite{LekCouVanDeb84}. All of these approaches gave similar results
for the properties of the coexisting isotropic and nematic phases.

While being able to solve explicitly only the monodisperse case,
Onsager~\cite{Onsager49} already outlined the possible extension of
the theory to {\em polydisperse} systems, \ie, to mixtures of rods of
different lengths and/or different diameters. Polydispersity has
indeed been recognized as an important feature affecting experimental
results~\cite{BuiLek93,VanVanLek96}, and some attempts have been made
to include it in theoretical
treatments~\cite{BirKolPry88,VroLek93,VroLek97,VanMul96}. A generic
prediction is a pronounced broadening of the coexistence region with
increasing
polydispersity~\cite{Evans01,Chen94,Sluckin89,ClaCueSeaSolSpe00},
which is also observed experimentally~\cite{VanVanLek96}. A second
generic effect of polydispersity is fractionation, \ie, the presence
of particles of different size in the coexisting phases; for rodlike
particles, already Onsager~\cite{Onsager49} had predicted that the
nematic phase would be enriched in the longer rods.  Polydispersity
can also result in more drastic and qualitative changes to the phase
behavior, however. In particular, in systems with length
polydispersity coexistence between two nematic phases (N-N) or one
isotropic and two nematic phases (I-N-N) can occur. This has been
observed experimentally~\cite{BuiLek93} and predicted theoretically
for bi- and tridisperse systems, \ie, mixtures of rods with two or
three different lengths~\cite{VroLek93,VroLek97}. However, a detailed
investigation of the effects of full length polydispersity, \ie, of a
continuous distribution of rod lengths, on the Onsager theory remains
an open problem. Perturbative approaches~\cite{Chen94,Sluckin89} by
their nature cannot access qualitative changes to the phase diagram
such as the occurrence of N-N or I-N-N coexistence. Our lead question
for this paper is therefore: can N-N and I-N-N coexistence occur in
length-polydisperse systems of thin hard rods? In cases where the
length distribution has two or three strong peaks, one expects
behavior similar to the bi- or tridisperse case, so that the answer
should be yes. Much less clear is what to expect for unimodal length
distributions, and this is the case that we will consider.

We concentrate on the onset of isotropic-nematic phase coexistence
coming from low density, \ie, on the isotropic cloud point; this can
be calculated numerically with some effort using an algorithm which we
have recently developed~\cite{SpeSol_bidisperse_ons}. Finding the
phase behavior for higher densities inside the coexistence region
would be substantially more difficult. We choose to start our analysis
from a fat-tailed (log-normal) length distribution with a finite upper
cutoff on rod length. This choice is inspired by the interesting
results obtained by \v{S}olc~\cite{Solc75,Solc70} for polydisperse
homopolymers, and by our recent investigation~\cite{SpeSol_p2_fat} of
length polydispersity 
effects within the $\Ptwo$ Onsager model; the latter is obtained by a
simplification of the angular dependence of the excluded volume of the
Onsager theory. We showed that within this
simplified model I-N-N coexistence is indeed possible in a system with
a log-normal (and hence unimodal) rod length distribution. The cloud
curve, which gives the
density where phase separation first occurs as a function of the width
of the length distribution, exhibits a kink where the system switches
between two different branches of I-N phase coexistence. The shadow
curve, which similarly records the density of the incipient nematic
``shadow'' phase, has a corresponding discontinuity. Precisely at the
kink in the cloud curve a single isotropic coexists with two different
nematics, so that this kink forms the beginning of an I-N-N
coexistence region. Both the cloud and the shadow curve were found to
depend strongly on the rod length cutoff; in the limit of large cutoff
the cloud and shadow curves approach the same limiting form, which is
universal for all length distributions with a fatter-than-exponential
tail. The nematic shadow phase has rather peculiar properties, being
essentially identical to the coexisting isotropic except for an
enrichment in the longest rods; the longer rods are also the only ones
that have significant orientational order.

The above results for the $\Ptwo$ Onsager model suggest that also in
the unapproximated Onsager theory a rod length distribution with a fat
tail should have pronounced effects on the phase behavior. We will
show numerically that the cloud curve indeed has a kink, and the shadow
curve a corresponding discontinuity, demonstrating that the phase
diagram contains a region of I-N-N coexistence. In fact, the effects
of the fat-tailed length distribution are even stronger than for the
$\Ptwo$ Onsager model, with the nematic shadow phase containing
essentially only the very longest rods in the system. The numerical
results leave open a number of questions, and we therefore supplement
it with a theoretical analysis. We show that the assumption of a
nematic shadow phase dominated by the longest rods is self-consistent,
and are able to predict that in the limit of large cutoff the density
of the cloud point actually tends to zero: even though the average rod
length is finite, the presence of a tail of long rods drives the
system to phase separate at any nonzero density.
%
%
Motivated by these results, we finally revisit the case of rod length
distributions with an exponential tail, using the Schulz distribution
as an example. Numerical results show, surprisingly, that even here a
regime occurs where the length cutoff matters and the nematic shadow
phase contains predominantly the longest rods; a kink in the cloud
curve again reveals the presence of an I-N-N coexistence region.
(This is stark contrast to our results for the $\Ptwo$ Onsager
model~\cite{SpeSol_p2}, where the exponential tail of the distribution
produces no unusual effects.) By returning to our theoretical analysis
we find that the long-rod effects are weaker for the Schulz
distribution than for the log-normal case: they only occur above a
certain threshold value for the width of the rod length distribution,
and the large-cutoff limits of the cloud and shadow densities above
this threshold remain nonzero.

The paper is structured as follows. In Sec.~\ref{sec:poly_ons} we
outline the extension to continuous length distributions of the
Onsager theory and derive the phase coexistence equations for the
isotropic cloud point. Sec.~\ref{sec:numerics}
begins with a brief description of the numerical method we used to
locate the cloud point; we then show our results for the phase
behavior for log-normal length distributions with finite cutoff. In
Sec.~\ref{sec:theory_lm} we outline our theory for the large cutoff
limit and compare with some numerical results at finite large cutoff,
finding good agreement. Finally, in Sec.~\ref{sec:schulz} we turn to
systems with a Schulz distribution of lengths, giving numerical
results and sketching an appropriately modified theoretical analysis
which again makes predictions in good agreement with the numerics. 
Sec.~\ref{sec:poly_ons_concl} contains a summary and a discussion of
avenues for future work. In
App.~\ref{app:high_dens_onsager} we review in outline the high density
scaling theory for the monodisperse Onsager theory which we exploit in
our analysis of the large cutoff limit. In
App.~\ref{app:long_rods_dom} the main approximation underlying our theory
for the log-normal distribution is justified, while the appropriate
modifications for the Schulz distribution are sketched in
App.~\ref{app:schulz_long}.

\section{The polydisperse Onsager theory}\label{sec:poly_ons}

The Onsager theory with length polydispersity models a system of hard
spherocylinders with equal diameters $D$ but different lengths $L$.
We introduce a reference lengthscale $L_0$ and write $L=lL_0$ where
$l$ is a dimensionless normalized length. The Onsager limit is then
taken by considering $D/L_0\rightarrow 0$ at constant values for the
normalized lengths $l$. From now on we will refer to $l$ itself as the
rod length unless stated otherwise; it can in principle range over
all values between $0$ and $\infty$.

The thermodynamic state of the system is described by the {\em density
distribution} $\rho(l,\Omega)$. This is defined such that 
$\rho(l,\Omega)\,dl\,(d\Omega/4\pi)$ is the number density of rods with
lengths in the range $l\ldots l+dl$ and pointing along a direction
within the solid angle $d\Omega$ around $\Omega$. In terms of
spherical coordinates, with the $z$-direction taken to be the nematic
axis, we have $d\Omega=\sin\theta\,d\theta\,d\varphi$ and
the density distribution is independent of the azimuthal angle $\varphi$,
$\rho(l,\Omega)\equiv\rhotl$. It can
thus be decomposed according to
\[
\rhotl=\rhol\ptl=\rho P(l)\ptl
\]
Here $\rhol$ is the density distribution over lengths,
\[
\rhol = \int\frac{d\Omega}{4\pi}\ \rhotl=\frac{1}{2}\int d\cos\theta\
\rhotl=\angint \rhotl
\]
where we have introduced the shorthand
\[
\thint=\frac{1}{2}d\cos\theta
\]
The overall rod number density is 
\[
\rho=\int dl\,\rhol = \totint\rtl
\]
so that $P(l)=\rho(l)/\rho$ gives the normalized length
distribution. From the definition of $\rhol$ it also follows that the
orientational distributions $\ptl$ for rods of fixed length are
normalized in the obvious way, $\angint \ptl=1$. Notice that the
factor $4\pi$ in the definition of $\rho(l,\Omega)$ has been chosen so
that for an isotropic phase one has the simple expressions $\ptl=1$
and $\rhol=\rtl$.

We can now state the free energy density for the polydisperse Onsager
theory (see Ref.~\onlinecite{Sluckin89}). We use units such that
$k_{\rm B}T=1$ and make all densities dimensionless by multiplying
with the unit volume $V_0=(\pi/4)DL_0^2$. The free energy density is
then
\bea\label{eq:ons_free_en}
f=\int dl\,
\rho(l)\left[\ln\rho(l)-1\right]+\totint\rho(l)\ptl\ln\ptl\nonumber\\
+\frac{1}{2}\int dl\, dl'\,\thint\,\thprint\, \rho(l)\rho(l')\ptl\ptlpr l
l'K(\theta,\theta')
\eea
The first term gives the entropy of an ideal mixture, while the second term
represents the orientational entropy of the rods. The third term
is the appropriate average of the excluded volume
$(8/\pi)V_0ll'|\sin\gamma|$ (with 
$V_0$ absorbed by our density scaling) of two rods at an angle
$\gamma$ with each other. The kernel~\cite{Onsager49,KayRav78}
$K(\theta,\theta')$ results from the average of $(8/\pi)|\sin\gamma|$ over
the azimuthal angles $\varphi, \varphi'$ of the rods,
\bea
K(\theta,\theta')&=&
\frac{8}{\pi}\int_0^{2\pi}\frac{d\varphi'}{2\pi}\frac{d\varphi}{2\pi}
|\sin\gamma|
\nonumber\\
%
%
&=&\frac{8}{\pi}\int_0^{2\pi}\frac{d\varphi}{2\pi}
\sqrt{1-
\left(\cos\theta\cos\theta'+\sin\theta\sin\theta'\cos\varphi\right)^2}
\nonumber
\eea
As in the monodisperse case, the orientational distributions $\ptl$
are obtained by minimization of the free
energy, Eq.~\eqref{eq:ons_free_en}; inserting Lagrange multipliers to
enforce the normalization of the $\ptl$, one finds
\bea
  \ptl&=&\dfrac{e^{l\psi(\theta)}}{\int\thprint\
    e^{l\psi(\theta')}}\label{eq:ons_ptl}\\ 
\psi(\theta)&=&-\int dl'\,
          \thprint\rho(l')\ptlpr l' K(\theta,\theta')\label{eq:psi_poly}
\eea
The conditions for phase equilibrium are that coexisting phases must have
equal chemical potential $\mu(l)$ for all rod lengths $l$, as well as
equal osmotic pressure. The chemical potentials can be obtained by
functional differentiation of the free energy~\eqref{eq:ons_free_en}
with respect to $\rhol$. The orientational distributions $\ptl$ do
depend on $\rhol$ but this dependence can be ignored because the
$\ptl$ are chosen to minimize $f$. Carrying out the differentiation
and inserting Eq.~\eqref{eq:ons_ptl} gives
\bea
\mu(l)&=&\frac{\delta f}{\delta\rhol}\nonumber\\
&=&\ln\rhol+\angint\ptl\left[l\psit-\ln\int\thprint\
e^{l\psitpr}\right]\nonumber\\
&&+\int dl'\,\thint\,\thprint\,
\rho(l')\ptl\ptlpr ll'K(\theta,\theta')\label{eq:ons_mu1}\\
&=&\ln\rhol-\ln\angint e^{l\psit}\label{eq:ons_mu}
\eea
The osmotic pressure can be obtained from the Gibbs-Duhem relation,
which for a polydisperse system reads
\[
\Pi=\int dl\ \mu(l)\rhol-f
\]
Inserting Eqs.~\eqref{eq:ons_ptl} and~\eqref{eq:ons_mu1} then yields
\beq\label{eq:ons_Pi}
\Pi=\rho-\frac{1}{2}\totint l\rhol\ptl\psit
\eeq


\subsection{Isotropic-nematic phase coexistence}\label{sec:I-N_eq}

We now specialize the phase coexistence conditions to I-N coexistence,
and then eventually to the isotropic cloud point, \ie, the onset of
I-N coexistence coming from low densities. The isotropic phase will
have
\beq
\iptl=1{\rm ,}\quad
\psi\I(\theta)=\psi\I=-c_1\rho\I_1{\rm ,}\quad
\Pi\I=\rho\I+\frac{1}{2}c_1\left(\rho\I_1\right)^2
\label{isotropic}
\eeq
where we have defined the first moment $\rho_1$ of the density
distribution $\rhotl$
\[
\rho_1=\int dl\, l\rhol=\rho\int dl\, l\pl=\rho\langle l\rangle
\]
which represents the scaled rod volume fraction, $\rho_1=(L_0/D)\phi$.
We have also used the fact that a uniform average of the kernel
$K(\theta,\theta')$ over one of its arguments is just an isotropic
average over $(8/\pi)|\sin\gamma|$, giving
\beastar
\int\thprint\,
K(\theta,\theta')&=&\frac{8}{\pi}\frac{1}{2}\int_0^\pi d\gamma\,
\sin\gamma |\sin\gamma|\\
&=&\frac{8}{\pi}\frac{\pi}{4}=2\equiv c_1
\eeastar
%
%
The equality of the chemical potentials~\eqref{eq:ons_mu} gives for
the density distribution in the nematic phase
\beq\label{eq:nematic}
\rho\N(l)=\rho\I(l)\angint e^{lg(\theta)}
\eeq
where 
\beq\label{eq:g_def}
\gt=\psi\N(\theta)-\psi\I=\psi\N(\theta)+c_1\rho\I_1
\eeq
The full density distribution over lengths and orientations is therefore,
using Eq.~\eqref{eq:ons_ptl},
\beq\label{eq:nem_phase}
\rho\N(l,\theta)=\rho\N\l P\N(\theta|l)=\rho\I(l)e^{l\gt}
\eeq
and the osmotic pressure~\eqref{eq:ons_Pi} of the nematic phase can
be rewritten as
\beq
\label{eq:n_Pi}
\Pi\N=\totint\rho\I(l)e^{l\gt}-\frac{1}{2}\totint
l\rho\I(l)e^{l\gt}\psi\N(\theta)
\eeq

In the following we will concentrate on the isotropic cloud point,
where the isotropic ``cloud'' phase starts to coexist with an
infinitesimal amount of nematic ``shadow'' phase. At the cloud point
the isotropic density distribution over lengths, $\rho\I(l)$,
therefore coincides with the overall density distribution of the
system, $\parent\l$, which we call the {\em parent distribution}. The
parent distribution can be written as $\parent(l)=\rho\normparent(l)$,
where $\rho=\int dl\,\parent(l)$ is the overall parent number density
and $\normparent(l)$ the normalized parent length distribution. Since
all properties of the isotropic cloud phase are determined by the
parent, we till drop the superscript ``I'' in the following. We will
also take the parent distribution to have average length $\langle
l\rangle=1$; any other choice could be absorbed into the reference
length $L_0$. This implies that the density and (scaled) volume
fraction of the isotropic phase are equal, $\rho_1=\rho$. With this
notation, the density distribution~\eqref{eq:nem_phase} of the nematic
shadow is $\rho\N(l,\theta)=\rho\normparent(l)e^{l\gt}$ and fully
determined by $\rho$ and $\gt$. The function $\gt$ must obey
\beq\label{eq:g_eq}
\gt=-\rho\int dl\,\thprint\,\normparent\l
e^{lg(\theta')}lK(\theta,\theta')+c_1\rho
\eeq
as follows from Eq.~\eqref{eq:psi_poly} for $\psi\N\t$ together with
Eq.~\eqref{eq:g_def}. The cloud point density is the smallest value of
$\rho$ for which in addition the pressure equality is satisfied. Using
Eq.~\eqref{isotropic} for the pressure of the isotropic and
Eqs.~\eqref{eq:g_def} and~\eqref{eq:n_Pi} for that of the nematic,
this condition reads
\bea
\rho+\frac{1}{2}c_1\rho^2&=&\rho\totint\normparent\l
e^{l\gt}\nonumber\\
&&{}-{}\frac{\rho}{2}\totint l\normparent\l
e^{l\gt}\left[\gt-c_1\rho\right]
\label{eq:Pi_eq}
\eea

\subsection{Fat-tailed rod length distributions}
\label{sec:fat_tail}

So far everything is general and applies to any parent length
distribution $\normparent\l$. Let us now focus on the case of a parent
distribution with a fat, \ie, less than exponentially decaying tail
for large $l$. At the cloud point we have from Eq.~\eqref{eq:nematic}
the density distribution $\rho\N\l=\rho\normparent\l\angint e^{l\gt}$
in the nematic shadow phase, or if we isolate the value of $\gt$ at
$\theta=0$
\beq\label{eq:fat_condition}
\rho\N\l=\rho\normparent\l e^{lg(0)}\angint e^{l[\gt-g(0)]}
\eeq
In a nematic, one expects $\gt\leq \gnot$ and therefore the angular
integral reduces to a less than exponentially varying function of
$l$. Moreover, $\gnot$ is expected to be positive, since the nematic
phase should contain the longer rods. The nematic density distribution
$\rho\N(l)$ is therefore exponentially diverging for large $l$
whenever the normalized parent distribution $\normparent(l)$ decays
less than exponentially. In order to ensure finite values for the
density and volume fraction of the nematic phase, we thus need to
impose a finite cutoff $l_m$ on the length distribution; unless
otherwise specified, all $l$-integrals will therefore run over
$0\ldots l_m$ from now on. The presence of a cutoff is of course also
physically reasonable, since any real system contains a finite largest
rod length. Nevertheless, we will later also consider the limit of
infinite cutoff, which highlights the effects of the presence of long
rods.

\section{Numerical results for the onset of I-N coexistence}
\label{sec:numerics}

\subsection{Numerical method}

A numerical determination of the isotropic cloud point involves the
solution of the two coupled equations~\eqq{eq:g_eq}{eq:Pi_eq} for
$\rho$ and $\gt$. In an outer loop we vary the density until the
smallest $\rho$ that satisfies the pressure equality~\eqref{eq:Pi_eq}
is found; we use
a false position method~\cite{isaacson}. The nontrivial part of the
algorithm is the inner loop, \ie, the solution of the functional
equation~\eqref{eq:g_eq} for $\gt$ at given $\rho$.  An iterative
method inspired by the one used by Herzfeld {\em et
al.}~\cite{HerBerWin84} for the monodisperse case turns out to
converge too slowly in the presence of polydispersity. We therefore
choose to represent $\gt$ by its values $g_i=g(\theta_i)$ at a set of
$n$ discrete points $\theta_i$; the values of $g(\theta)$ for
$\theta\neq\theta_i$ are then assumed to be given by a cubic spline
fit~\cite{NumRec} through the points $(\theta_i,g_i)$. This turns the
functional Eq.~\eqref{eq:g_eq} into a set of $n$ nonlinear coupled
equations
%
%
which can be solved by \eg\ a Newton-Raphson
algorithm~\cite{isaacson}.  To keep $n$ manageably small while keeping
the spline representation accurate, a judicious choice of the
$\theta_i$ is important. We exploit the symmetry $\gt=g(\pi-\theta)$
and choose a nonlinear (geometric) spacing of the $\theta_i$ over the
range $0\ldots\pi/2$, with more points around the origin where $g\t$
is least smooth.

The numerical method outlined above was also used in our recent
investigation of the exact phase diagram for the bidisperse
Onsager theoery~\cite{SpeSol_bidisperse_ons}. The only modification is
in the $l$-integrations, which in the bidisperse case became
simple sums over the two rod lengths. In principle the additional
integration could make the calculation substantially slower; however,
one notices that in Eqs.~\eqref{eq:g_eq} and~\eqref{eq:Pi_eq} only the
two $l$-integrals
\[
h(x)=\int dl\ \normparent(l)\, e^{lx}
\]
and
\[
h'(x)=\int dl\ \normparent(l)\, le^{lx}
\]
are needed. We therefore precompute these functions and store them
once and for all as cubic spline fits which can be evaluated very
efficiently.

As an alternative to the approach above we also considered calculating
$g(\theta)$ by minimizing an appropriate functional. If $\mu\I(l)$ and
$\Pi\I$ are the chemical potentials and osmotic pressure of the
isotropic parent phase, then one easily sees that a local minimum of
the functional
\beq\label{eq:def_G}
\Xi[\rhotl]=f[\rhotl]-\totint \mu\I(l)\rhotl + \Pi\I
\eeq
corresponds to a phase $\rhotl$ with the same chemical potentials as
the parent. Inserting for $\rhotl$ the known form~\eqref{eq:nem_phase}
for the nematic density distribution, $\Xi$ turns into a functional of
$g(\theta)$, and the condition for a local minimum becomes equivalent
to Eq.~\eqref{eq:g_eq}. $\Xi[g(\theta)]$ always has $g(\theta)\equiv
0$ as a minimum, corresponding to the isotropic parent itself; but
close to the onset of phase coexistence an additional nematic solution with
$g(\theta)\neq 0$ appears. Notice that, geometrically, $\Xi$ is a tilted
version of the free energy for which the tangent (hyper-)plane at the
parent is horizontal ($\Xi=0$). At phase coexistence, \ie, for a parent
with the cloud point density, the tangent plane also touches the
nematic and so the isotropic and the nematic minimum of $\Xi$ are
both at ``height'' $\Xi=0$.

Numerically, one could minimize $\Xi$ by again representing
$g(\theta)$ as a spline through a finite number of points
$g_i=g(\theta_i)$ and then minimizing the resulting function of the
$g_i$.  In general this turns out to be no easier than the solution of
the similarly discretized version of Eq.~\eqref{eq:g_eq}. However, the
minimization approach is useful when $\gt$ assumes a simple parametric
form. We will see later that this is indeed the case for large cutoff
$l_m$, with $\gt$ being well approximated by the two-parameter form
$g\t=a-b\sin\theta$. Inserting this into $\Xi$ and minimizing over $a$
and $b$ then gives an approximate solution for $\gt$; when used as a
starting point for $\gt$, this makes it significantly easier to
converge the numerical solution of the discretized
Eq.~\eqref{eq:g_eq} described above.

\subsection{Results for log-normal length distribution}

With the numerical method described in the previous section, it is
possible to solve for the onset of isotropic-nematic phase
coexistence for in principle arbitrary parent length distribution. We
choose here a specific fat-tailed length distribution, the log-normal,
which has already given interesting results in polymers~\cite{Solc75}
and in our previous analysis of the $\Ptwo$ Onsager
model~\cite{SpeSol_p2_fat}. The log-normal distribution has the form
\beq\label{eq:log-normal_ons} \normparent(l)=\frac{1}{\sqrt{2\pi
w^2}}\frac{1}{l}\exp\left[-\frac{(\ln l-\mu)^2}{2 w^2}\right]
\eeq
with finite length cutoff $l_m$. The quantity $w$, which tunes the
width of the distribution, is fixed by the normalized standard
deviation $\sigma$ (in the following referred to as polydispersity)
\[
\sigma^2=\frac{\langle l^2\rangle-\langle l\rangle ^2}{\langle l\rangle^2}
\]
to be $w^2=\ln(1+\sigma^2)$. The second parameter $\mu$ is determined
so that the parent has average length $\langle l\rangle=1$, giving
$\mu=-w/2$. Notice that with these choices, the parent length
distribution is normalized and has the desired moments $\langle
l\rangle=1$ and $\langle l^2\rangle=1+\sigma^2$ only in the limit of
infinite cutoff $l_m$. The deviations for finite cutoffs are small
even for relatively modest $l_m$, however. For instance, at cutoff
$l_m=50$ and $\sigma=0.5$, the integrals $\int_0^{l_m} dl\
l^n\normparent(l)$ for $n=0,1,2$ differ from their respective values
at infinite cutoff, 1, 1 and 1.25, by values of order $10^{-17}$,
$10^{-15}$ and $10^{-13}$. Since we will not consider smaller cutoff
values below, these small corrections can safely be neglected.

In our previous analysis of the $\Ptwo$ Onsager
model~\cite{SpeSol_p2_fat}, we observed that for log-normal length
distributions the cloud curve has a kink, and the shadow curve a
corresponding discontinuity; at the kink, the isotropic phase is in
coexistence with two distinct nematics, so that the phase diagram must
contain a region of I-N-N coexistence. In the $\Ptwo$ Onsager case,
the simplicity of the model actually allowed us to compute the
complete phase diagram and locate the three-phase I-N-N region
explicitly. For the full Onsager theory treated here we can only find
the cloud and shadow curves at present, not the full phase diagram;
nevertheless, a kink in the cloud curve will again imply the presence
of an I-N-N region in the full phase diagram.
\begin{figure}[htb]
\begin{center}
\begin{picture}(0,0)%
\includegraphics{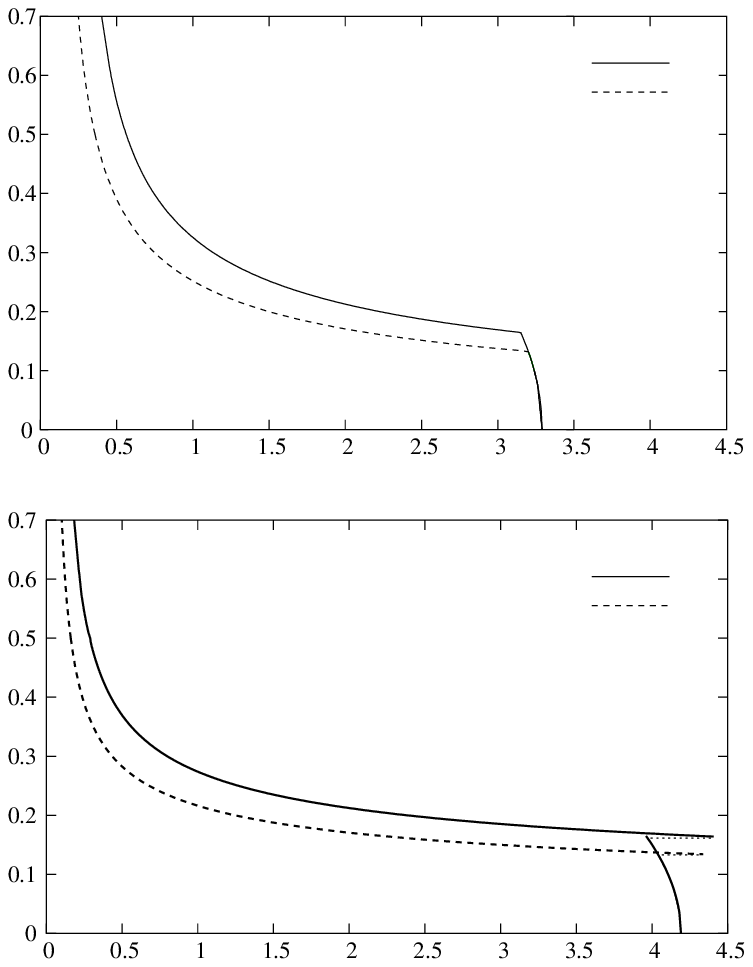}%
\end{picture}%
\setlength{\unitlength}{2447sp}%
\begingroup\makeatletter\ifx\SetFigFont\undefined
\def\x#1#2#3#4#5#6#7\relax{\def\x{#1#2#3#4#5#6}}%
\expandafter\x\fmtname xxxxxx\relax \def\y{splain}%
\ifx\x\y   
\gdef\SetFigFont#1#2#3{%
  \ifnum #1<17\tiny\else \ifnum #1<20\small\else
  \ifnum #1<24\normalsize\else \ifnum #1<29\large\else
  \ifnum #1<34\Large\else \ifnum #1<41\LARGE\else
     \huge\fi\fi\fi\fi\fi\fi
  \csname #3\endcsname}%
\else
\gdef\SetFigFont#1#2#3{\begingroup
  \count@#1\relax \ifnum 25<\count@\count@25\fi
  \def\x{\endgroup\@setsize\SetFigFont{#2pt}}%
  \expandafter\x
    \csname \romannumeral\the\count@ pt\expandafter\endcsname
    \csname @\romannumeral\the\count@ pt\endcsname
  \csname #3\endcsname}%
\fi
\fi\endgroup
\begin{picture}(6075,7688)(1051,-8135)
\put(1051,-1861){\makebox(0,0)[lb]{\smash{\SetFigFont{7}{8.4}{rm}{\color[rgb]{0,0,0}$\sigma$}%
}}}
\put(1051,-5761){\makebox(0,0)[lb]{\smash{\SetFigFont{7}{8.4}{rm}{\color[rgb]{0,0,0}$\sigma$}%
}}}
\put(4276,-8086){\makebox(0,0)[lb]{\smash{\SetFigFont{7}{8.4}{rm}{\color[rgb]{0,0,0}$\rho\N$}%
}}}
\put(4294,-4141){\makebox(0,0)[lb]{\smash{\SetFigFont{7}{8.4}{rm}{\color[rgb]{0,0,0}$\rho$}%
}}}
\put(5001,-891){\makebox(0,0)[lb]{\smash{\SetFigFont{7}{8.4}{rm}{\color[rgb]{0,0,0}$l_m=50$}%
}}}
\put(5001,-1111){\makebox(0,0)[lb]{\smash{\SetFigFont{7}{8.4}{rm}{\color[rgb]{0,0,0}$l_m=100$}%
}}}
\put(5001,-4881){\makebox(0,0)[lb]{\smash{\SetFigFont{7}{8.4}{rm}{\color[rgb]{0,0,0}$l_m=50$}%
}}}
\put(5001,-5101){\makebox(0,0)[lb]{\smash{\SetFigFont{7}{8.4}{rm}{\color[rgb]{0,0,0}$l_m=100$}%
}}}
\put(6501,-2256){\makebox(0,0)[lb]{\smash{\SetFigFont{7}{8.4}{rm}{\color[rgb]{0,0,0}(a)}%
}}}
\put(6501,-6061){\makebox(0,0)[lb]{\smash{\SetFigFont{7}{8.4}{rm}{\color[rgb]{0,0,0}(b)}%
}}}
\end{picture}
\vspace*{0.3cm}
\caption{(a) Number density $\rho$ of the cloud phase for length
cutoff $l_m=50$ (solid) and $l_m=100$ (dashed), plotted against the
polydispersity $\sigma$ on the $y$-axis. Notice the kinks in the two
curves, which imply the presence of a three-phase I-N-N coexistence
region in the full phase diagram. The kinks correspond, as they
should, to discontinuities in the shadow curves (b). Both cloud and
shadow curves are strongly cutoff-dependent, moving towards lower
densities as $l_m$ increases.  }
\label{fig:cloud_shad_50_100}
\efig
In Fig.~\ref{fig:cloud_shad_50_100} we show the cloud and shadow
curves, \ie, the number density of the isotropic cloud (a) and nematic
shadow (b) plotted against polydispersity, for a log-normal parent
with two different cutoffs. The presence of the kink in the cloud
curves, and the corresponding discontinuity in the shadow curves, is
clear evidence of the presence of an I-N-N coexistence region starting
at the kink of the cloud curve. The positions of the kink and
discontinuity, respectively, as well as the shapes of the cloud and
shadow curves above them, show a strong dependence on the cutoff
length; both curves move to significantly smaller densities as $l_m$
increases. For polydispersities $\sigma$ below the kink, on the other
hand, the number of long rods is too small to have a significant
effect on the phase separation and one has essentially
cutoff-independent behavior that connects smoothly with the
monodisperse limit $\sigma=0$. These observations are in qualitative
accord with our earlier results for the $\Ptwo$ Onsager model with the
same length distribution~\cite{SpeSol_p2_fat}.

Moving across the discontinuity from below, the shadow curve jumps
from a ``normal'' nematic branch to an unusual nematic phase which, as
we will see below, is completely dominated by the longest rods in the
parent distribution.
\begin{figure}[htb]
\begin{center}
\begin{picture}(0,0)%
\includegraphics{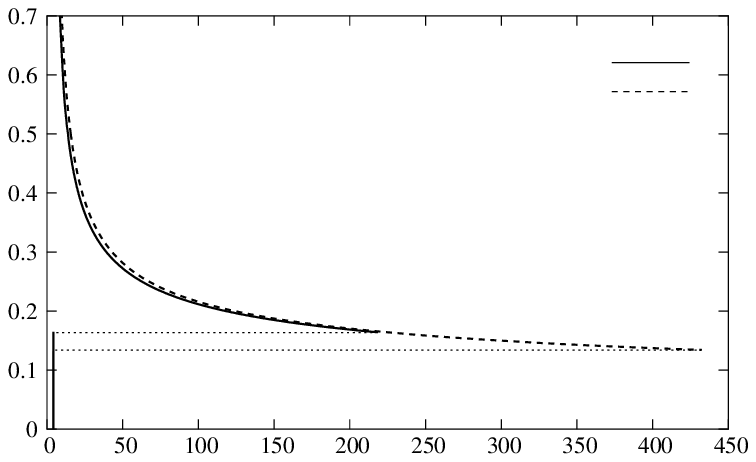}%
\end{picture}%
\setlength{\unitlength}{2447sp}%
\begingroup\makeatletter\ifx\SetFigFont\undefined
\def\x#1#2#3#4#5#6#7\relax{\def\x{#1#2#3#4#5#6}}%
\expandafter\x\fmtname xxxxxx\relax \def\y{splain}%
\ifx\x\y   
\gdef\SetFigFont#1#2#3{%
  \ifnum #1<17\tiny\else \ifnum #1<20\small\else
  \ifnum #1<24\normalsize\else \ifnum #1<29\large\else
  \ifnum #1<34\Large\else \ifnum #1<41\LARGE\else
     \huge\fi\fi\fi\fi\fi\fi
  \csname #3\endcsname}%
\else
\gdef\SetFigFont#1#2#3{\begingroup
  \count@#1\relax \ifnum 25<\count@\count@25\fi
  \def\x{\endgroup\@setsize\SetFigFont{#2pt}}%
  \expandafter\x
    \csname \romannumeral\the\count@ pt\expandafter\endcsname
    \csname @\romannumeral\the\count@ pt\endcsname
  \csname #3\endcsname}%
\fi
\fi\endgroup
\begin{picture}(6028,3788)(1126,-4235)
\put(5211,-1111){\makebox(0,0)[lb]{\smash{\SetFigFont{7}{8.4}{rm}{\color[rgb]{0,0,0}$l_m=100$}%
}}}
\put(5211,-891){\makebox(0,0)[lb]{\smash{\SetFigFont{7}{8.4}{rm}{\color[rgb]{0,0,0}$l_m=50$}%
}}}
\put(4501,-4186){\makebox(0,0)[lb]{\smash{\SetFigFont{7}{8.4}{rm}{\color[rgb]{0,0,0}$\rho\N_1$}%
}}}
\put(1126,-2011){\makebox(0,0)[lb]{\smash{\SetFigFont{7}{8.4}{rm}{\color[rgb]{0,0,0}$\sigma$}%
}}}
\end{picture}
\vspace*{0.3cm}
\caption{Scaled volume fraction $\rho\N_1$ of the nematic shadow phase at
  $l_m=50$ (solid) and $l_m=100$ (dashed). Notice that the
  discontinuity of the two curves is now much wider than in the number
  density representation of the shadow curves in
  Fig.~\protect\ref{fig:cloud_shad_50_100}(b). 
}
\label{fig:rho1_50_100}
\efig
In the rescaled volume fraction ($\rho_1$) representation of the
shadow curve shown in Fig.~\ref{fig:rho1_50_100} the different
characteristics of the two nematic phases are clear. While at
low polydispersity $\rho\N_1$ is of order unity, it jumps by two
orders of magnitude on crossing the discontinuity. This shows that the
fractionation effect which one normally encounters in polydisperse
systems, with the long rods found preferentially in the nematic phase,
becomes extreme here. A plot of the average rod length in the nematic
shadow against polydispersity (Fig.~\ref{fig:length_50_100}) in fact
shows that above the discontinuity, the nematic phase contains almost
exclusively rods of length close to the cutoff length.
\begin{figure}[htb]
\begin{center}
\begin{picture}(0,0)%
\includegraphics{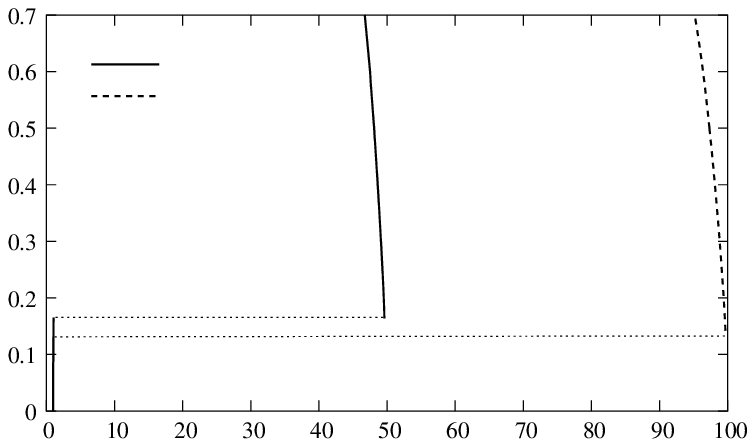}%
\end{picture}%
\setlength{\unitlength}{2447sp}%
\begingroup\makeatletter\ifx\SetFigFont\undefined
\def\x#1#2#3#4#5#6#7\relax{\def\x{#1#2#3#4#5#6}}%
\expandafter\x\fmtname xxxxxx\relax \def\y{splain}%
\ifx\x\y   
\gdef\SetFigFont#1#2#3{%
  \ifnum #1<17\tiny\else \ifnum #1<20\small\else
  \ifnum #1<24\normalsize\else \ifnum #1<29\large\else
  \ifnum #1<34\Large\else \ifnum #1<41\LARGE\else
     \huge\fi\fi\fi\fi\fi\fi
  \csname #3\endcsname}%
\else
\gdef\SetFigFont#1#2#3{\begingroup
  \count@#1\relax \ifnum 25<\count@\count@25\fi
  \def\x{\endgroup\@setsize\SetFigFont{#2pt}}%
  \expandafter\x
    \csname \romannumeral\the\count@ pt\expandafter\endcsname
    \csname @\romannumeral\the\count@ pt\endcsname
  \csname #3\endcsname}%
\fi
\fi\endgroup
\begin{picture}(6093,3674)(1051,-4236)
\put(2736,-1315){\makebox(0,0)[lb]{\smash{\SetFigFont{7}{8.4}{rm}{\color[rgb]{0,0,0}$l_m=100$}%
}}}
\put(2736,-1068){\makebox(0,0)[lb]{\smash{\SetFigFont{7}{8.4}{rm}{\color[rgb]{0,0,0}$l_m=50$}%
}}}
\put(1051,-2011){\makebox(0,0)[lb]{\smash{\SetFigFont{7}{8.4}{rm}{\color[rgb]{0,0,0}$\sigma$}%
}}}
\put(4351,-4186){\makebox(0,0)[lb]{\smash{\SetFigFont{7}{8.4}{rm}{\color[rgb]{0,0,0}$\langle l\rangle\N$}%
}}}
\end{picture}
\vspace*{0.3cm}
\caption{Average length $\langle l\rangle\N=\rho_1\N/\rho\N$ in the nematic
  shadow phase against polydispersity, for cutoffs $l_m=50$ (solid)
  and $l_m=100$ (dashed). Above the discontinuity, fractionation
  becomes extreme. In fact, at the discontinuity the nematic phase
  appears to be composed only of rods with lengths of order the cutoff
  length. At the top of the plot, for very large polydispersities
  where the parent length distribution becomes somewhat more uniform,
  some shorter rods are also included in the nematic phase and reduce
  the average length.
}
\label{fig:length_50_100}
\efig
%
%
Since the longer rods are generically expected to be
more strongly ordered, one should then also
find that the nematic shadow phase has very strong orientational
order. This is indeed the case: the orientational order parameter $S$,
defined as an average in the nematic phase over the second Legendre
polynomial $\Ptwo(\cos\theta)=(3\cos^2\theta-1)/2$,
\[
S=\totint P\N(l)P\N(\theta|l)\Ptwo(\cos\theta)
\]
is almost indistinguishable from 1 above the discontinuity in the
nematic shadow curve, implying that the typical angles $\theta$ which
the rods make with the nematic axis are very small. This also implies
that the rods we consider must be rather thin for Onsager's second
virial approximation to be valid: for monodisperse
rods~\cite{VroLek92}, the criterion is $D/L\ll\theta$. We return to
this point in Sec.~\ref{sec:ons_validity}.

The results shown above for the finite cutoff regime leave open a
number of questions.  For example, we observed that both the isotropic
cloud and nematic shadow curves move to lower densities as the cutoff
increases, but the modest $l_m$-values used are too small to determine
whether the curves will converge to a nonzero limit as $l_m$ grows
large or instead approach zero. One would also like to ascertain
whether the average rod length in the nematic shadow really tends to
$l_m$ for large cutoffs, as suggested by Fig.~\ref{fig:length_50_100},
and what happens to the rescaled nematic volume fraction $\rho\N_1$ in
the limit (Fig.~\ref{fig:rho1_50_100} suggests that it might become
large). In the next section we therefore turn to a theoretical
analysis of the limit $l_m\to\infty$, which will clarify all these
points.

\section{Theory for fat-tailed distributions with large cutoff}
\label{sec:theory_lm}

Above we saw that, at the onset of isotropic-nematic phase coexistence
in systems with log-normal length distributions, the nematic shadow
phase appears to be dominated by the longest rods in the distribution,
with lengths $l\approx l_m$. In this section we will construct a
consistent theory for the phase behavior in the limit of large
cutoffs $l_m$ based on this hypothesis.  From this we will be able to
extract the limiting dependence on $l_m$ of the cloud point density
$\rho$, the density $\rho\N$ and (scaled) volume fraction $\rho\N_1$ of the
nematic shadow phase, and the function $g\t$ determining the
orientational ordering of the nematic phase. At the end of this
section we will then compare these predictions with 
numerical results obtained for finite but large cutoff.

\subsection{Dominance of long rods in the nematic}

Recall that, in principle, we need to solve Eqs.~\eqref{eq:g_eq}
and~\eqref{eq:Pi_eq} for $\gt$ and $\rho$ to determine the isotropic
cloud point and the properties of the coexisting nematic shadow
phase. It will be useful to isolate, in Eq.~\eqref{eq:g_eq}, the
contribution $e^{l\gnot}$ which determines the divergence of the
nematic density distribution~\eqref{eq:fat_condition}. Define the
function
\beq
\label{eq:h_poly_def}
h\t=l_m\left[\gnot-g\t\right]
\eeq 
with the sign chosen so that $h\t$ should be
non-negative for all $\theta$. Eq.~\eqref{eq:g_eq} evaluated at
$\theta=0$ then gives
\bea
\gnot&=&-\rho\int dl\ \normparent\l e^{l\gnot} l \int\thprint\ e^{
-(l/l_m)h(\theta')} K(0,\theta')
\nonumber\\
& &{}+{}c_1\rho
\label{eq:gnot1}
\eea 
and subtracting from Eq.~\eqref{eq:g_eq} yields
\bea
h\t&=&\rho l_m\int dl\ \normparent\l e^{l\gnot} l\nonumber\\
& &{}\times{}\int\thprint\
e^{-({l}/{l_m})h(\theta')}\left[K(\theta,\theta')-K(0,\theta')\right]
\label{eq:h_theta1}
\eea
Together, Eqs.~\eqref{eq:gnot1} and~\eqref{eq:h_theta1} for $g(0)$ and
$h(\theta)$ are of course equivalent to Eq.~\eqref{eq:g_eq} for $g\t$.

To formalize the assumption that the nematic phase is dominated by the
longest rods, consider now the nematic density
distribution~\eqref{eq:fat_condition} which in our new notation reads
\beq
\rho\N\l=\rho\normparent\l e^{lg(0)}\angint e^{-(l/l_m)h\t}
\label{eq:rhoNl_new}
\eeq
If the exponential factor $\exp[lg(0)]$ is large enough for the
nematic indeed to be dominated by the longest rods, we can replace the
weakly varying (at most as a power law in $l$) angular integral by its
value at $l=l_m$, giving for the nematic density $\rho\N=\int dl\
\rho\N(l)$
\beq
\rho\N= \rho\int dl\ \normparent(l)e^{l\gnot}\angint e^{-h\t}
\label{eq:rhoN_first}
\eeq
We can now make the same approximation in Eq.~\eqref{eq:gnot1}: the
weakly varying factor in the $l$-integral is $l$ times the angular
integral, and replacing this by its value at $l=l_m$ yields
\beq\label{eq:gnot}
\gnot = -\rho l_m\int dl\ \normparent\l e^{l\gnot} \int\thint\
e^{ -h\t } K(0,\theta) + c_1\rho
\eeq 
In Eq.~\eqref{eq:h_theta1} for $h\t$, the analogous approximation gives
\bea
h\t &=&  \rho l_m^2 \int dl\ \normparent(l)e^{l\gnot}
\nonumber\\
& &{}\times{}\int\thprint\
e^{-h(\theta')}\left[K(\theta,\theta')-K(0,\theta')\right]
\label{eq:h_theta}
\eea 
Finally, consider the expression on the r.h.s of Eq.~\eqref{eq:Pi_eq} for
the osmotic pressure $\Pi\N$ in the nematic. The first term is the
ideal contribution, \ie, the nematic density $\rho\N$, while inserting
the definition~\eqref{eq:h_poly_def} in the second term gives
\bea
\Pi\N &=& \rho\N - 
\frac{\rho}{2}\lint \normparent\l e^{l\gnot} l\nonumber\\ 
&\times&\angint e^{-(l/l_m) h\t} \left[-\frac{1}{l_m}h\t+\gnot-c_1\rho\right]
\label{eq:Pi_first}
\eea
Replacing $l$ by $l_m$ in the weakly varying terms then yields
\bea
\Pi\N &=& \rho\N - 
\frac{\rho}{2}\lint \normparent\l e^{l\gnot} \nonumber\\
&\times&\angint e^{-h\t} \left[-h\t+l_m(\gnot-c_1\rho)\right]
\label{eq:Pi_second}
\label{eq:Pi_approx}
\eea
We show in App.~\ref{app:long_rods_dom} that a posteriori, the
approximation of dominance of the long rods can be justified in all
cases above, with the contributions to the $l$-integrals from rod
lengths $l\ll l_m$ becoming negligible for $l_m\to\infty$.

\subsection{The large-cutoff scaling solution}

We have now got four equations to be solved for $\rho\N$, $g(0)$,
$h\t$ and $\rho$, appropriately simplified using the assumption that
the nematic phase is dominated by the longest rods in the parent
distribution. These are Eqs.~\eqref{eq:rhoN_first}, \eqref{eq:gnot}
and~\eqref{eq:h_theta} and the pressure equality
$\Pi\N=\rho+(c_1/2)\rho^2$ with $\Pi\N$ from Eq.~\eqref{eq:Pi_second}.
Using Eq.~\eqref{eq:rhoN_first}, Eq.~\eqref{eq:h_theta} can be written as
\beq\label{eq:poly_h}
h\t=\rhoeff\frac{\int\thprint\
  e^{-h\tpr}\left[K(\theta,\theta')-K(0,\theta')\right]}{\int
  \widetilde{d\theta'}\ e^{-h(\theta')}}
\eeq
Here we have defined
\beq\label{eq:rhoeff_def}
\rhoeff=\rho\N l_m^2
\eeq
which is just the dimensionless density of the nematic phase, with the
factor $l_m^2$ arising from the fact that, since the nematic is
effectively monodisperse with $l=l_m$, one should use $l_mL_0$ rather
than $L_0$ in the definition of the unit volume $V_0$ (see before
Eq.~\eqref{eq:ons_free_en}). In the form
above, Eq.~\eqref{eq:poly_h} is identical to
Eq.~\eqref{eq:mono_ons_h} in App.~\ref{app:high_dens_onsager} for a
{\em monodisperse} system at dimensionless density
$\rhoeff$. Anticipating that $\rhoeff$ will be large, an assumption to
be checked a posteriori, we then deduce immediately that $h\t$ will be
given by the high-density scaling solution sketched in
App.~\ref{app:high_dens_onsager}, $h\t = \hs(t)$, with the scaling
variable $t=\rhoeff\sin\theta$.  The scaling function is determined by
Eq.~\eqref{eq:scaling_eq_mono}
\beq\label{eq:scaling_h_poly}
\hs(t)= \langle \Ks(t,t')-\Ks(0,t') \rangle_{t'}
\eeq
where the average over $t'$ is over the normalized distribution 
\beq
\Ps(t) =\frac{1}{\gamma}\, t\, e^{-\hs(t)}, \qquad
\gamma=\int dt\ t\,e^{-\hs(t)}
\label{eq:gamma_norm}
\eeq
and $\Ks(t,t')$ is the small-angle scaling form of the kernel
$K(\theta,\theta')$ given in Eq.~\eqref{eq:Ks_def} in
App.~\ref{app:high_dens_onsager}. The range of all averages and
integrals over $t$ and $t'$ can be taken as $0 \ldots \infty$ (rather
than $0\ldots \rhoeff$) for large $\rhoeff$ since the large-$t$ regime
gives only a negligible contribution.

With the form of $h\t$ determined up to the single parameter
$\rhoeff=\rho\N l_m^2$, just three unknowns $\rho$, $\rho\N$ and
$\gnot$ now remain, to be determined from Eq.~\eqref{eq:rhoN_first},
Eq.~\eqref{eq:gnot} and the osmotic pressure equality. We now simplify
these relations further, making use of the fact that in all angular
integrals involving the factor
$\exp[-h(\theta)]=\exp[-\hs(\rhoeff\sin\theta)]$ only small angles
$\theta \sim 1/\rhoeff$ contribute significantly.  Physically, this
means that the rods in the nematic shadow phase, which is at high
dimensionless density $\rhoeff$, have strong orientational order. For
such small $\theta$ we can set $\sin\theta\approx \theta$ and
transform everywhere to the scaling variable
$t=\rhoeff\sin\theta\approx \rhoeff\theta$. Eq.~\eqref{eq:rhoN_first}
then becomes, using the definition~\eqref{eq:gamma_norm}
\bea
\rho\N &=& \rho\int dl\
\normparent(l)e^{l\gnot}\int \frac{dt\ t}{\rhoeff^2}\,e^{-\hs(t)}\nonumber\\
&=& \frac{\rho\gamma}{\rhoeff^2} \int dl\
\normparent(l)e^{l\gnot}\label{eq:rhoN_rho}
\eea
Eq.~\eqref{eq:gnot} can be similarly transformed and, using
$K(0,\theta)=(8/\pi)\sin\theta=(8/\pi)t/\rhoeff$, reads
\bea
\gnot&=& - \rho l_m
\int dl\ \normparent\l e^{l\gnot} \int \frac{dt\ t}{\rhoeff^2}\,
e^{-\hs(t)}\left(\frac{8}{\pi}\frac{t}{\rhoeff}\right)
\nonumber\\
& &{}+{}c_1\rho
\label{eq:g0_first}
\eea
Comparing with Eq.~\eqref{eq:rhoN_rho}, and using $\rho\N
l_m/\rhoeff=1/l_m$, this can be written in simpler form as an average
over the distribution~\eqref{eq:gamma_norm},
\beq\label{eq:gnot_asy}
\gnot=
-\frac{8}{\pi}\frac{\langle t\rangle_t}{l_m}+c_1\rho
\eeq
Finally, the expression~\eqref{eq:Pi_second} for the osmotic pressure
in the nematic can also be simplified by using that $\theta$ is small
and transforming to the scaling variable $t$. Inserting
Eq.~\eqref{eq:gnot_asy}, this gives
\bea
\Pi\N&=&\rho\N+\frac{\rho}{2}\int dl\ \normparent(l)e^{l\gnot}
\int \frac{dt\ t}{\rhoeff^2}\,e^{-\hs(t)}
\left[\hs(t)+\frac{8}{\pi}\langle
  t\rangle_t\right]
\nonumber\\
&=& \rho\N+\frac{\rho}{2}\int dl\
\normparent(l)e^{l\gnot}\frac{\gamma}{\rhoeff^2}\left[\langle\hs(t)\rangle_t+\frac{8}{\pi}\langle t\rangle_t\right]\nonumber
\eea
In App.~\ref{app:high_dens_onsager} we show
(Eq.~\eqref{eq:scaling_excl_vol_mono}) that the scaling properties of
the high-density limit imply that the constant in the square brackets
has the value $4$; using also Eq.~\eqref{eq:rhoN_rho} we then get the
simple result
\beq
\Pi\N=3\rho\N
\label{eq:Pi_scaling_poly}
\eeq
This is identical to Eq.~\eqref{eq:Pi_linear} derived in
App.~\ref{app:high_dens_onsager} for monodisperse nematics at high
(dimensionless) densities, and therefore consistent with our
assumption that the nematic shadow phase behaves as an effectively
monodisperse system. The isotropic parent phase has pressure
$\Pi\I=\rho+(c_1/2)\rho^2$, so that the pressure equality
takes the form
\beq\label{eq:Pi_eq_asy}
\rho\N=\frac{1}{3}\left(\rho+\frac{c_1}{2}\rho^2\right)
\eeq

We can now proceed to determine the $l_m$-dependence of $\rho$,
$\rho\N$ and $g(0)$. Multiplying Eq.~\eqref{eq:rhoN_rho} by $(\rho\N)^2$
and inserting Eq.~\eqref{eq:Pi_eq_asy} gives
\[
\frac{\rho^3}{27}\left(1+\frac{c_1}{2}\rho\right)^3
=\rho\frac{\gamma}{l_m^4}\int dl\ \normparent(l)e^{l\gnot}
\]
The $l$-integral will again be dominated by the longest rods, so that
we can set $\normparent(l)=\normparent(l_m)$ to leading order
(see Ref.~\onlinecite{SpeSol_p2_fat}) to obtain
\[
\frac{\rho^2}{27}\left(1+\frac{c_1}{2}\rho\right)^3
=\frac{\gamma}{l_m^4}\normparent(l_m)\frac{e^{l_m\gnot}}{\gnot}
\]
Inserting Eq.~\eqref{eq:gnot_asy} to eliminate $\gnot$, we finally get
a nonlinear equation relating $\rho$ and $l_m$
\[
e^{l_mc_1\rho}=
\frac{\rho^2l_m^4}{27\gamma}\left(1+\frac{c_1}{2}\rho\right)^3\left(c_1\rho-\frac{8}{\pi}\frac{\langle
    t\rangle_t}{l_m}\right)\frac{e^{(8/\pi)\langle
t\rangle}}{\normparent(l_m)}
\] 
To obtain the asymptotic solution for large $l_m$, we anticipate that
$\rho$ will vary no stronger than a power law with $l_m$; for all
fat-tailed parent distributions except those with power-law tails, the
dominant $l_m$-dependence on the r.h.s.\ will then be through the
factor $1/\normparent(l_m)$.  Taking logs we have
\beq
l_m c_1\rho=-\ln\normparent(l_m)+\order(\ln l_m)
\eeq
Specializing to log-normal parent distributions, with
$\ln\normparent(l) = -(\ln^2 l)/(2w^2)$ to leading order, we
finally get
\beq\label{eq:rho_scaling}
\rho=\frac{\ln^2l_m}{2c_1 w^2 l_m}+\order\left(\frac{\ln
    l_m}{l_m}\right)
\eeq
showing that $\rho$ indeed varies with $l_m$ as a power-law (with
logarithmic corrections). Our theory thus predicts that the isotropic
cloud point density converges to zero for large cutoffs; in the
extreme limit of a log-normal parent distribution with an infinite
cutoff, phase separation would occur at any nonzero density.  From
Eq.~\eqref{eq:Pi_eq_asy}, the density of the nematic shadow phase
likewise vanishes for large cutoff, with $\rho\N=\rho/3$ to
leading order. For $g(0)$ we have from Eq.~\eqref{eq:gnot_asy}
\beq\label{eq:gnot_scaling}
\gnot=\frac{\ln^2l_m}{2w^2 l_m}+\order\left(\frac{\ln
l_m}{l_m}\right) 
- \frac{8}{\pi}\frac{\langle t\rangle_t}{l_m}
\eeq
and the last term of $\order(1/l_m)$ is subdominant compared to both the
first term and the first order correction $\order((\ln
l_m)/l_m)$. We can also now write down the whole function $g\t$, using
$g\t = g(0)-h\t/l_m = g(0)-\hs(\rho\N l_m^2 \sin\theta)/l_m$.  From
App.~\ref{app:high_dens_onsager} we know that the scaling function
$\hs$ is given by $\hs(t)=(8/\pi)t$ up to correction terms of
$\order(1)$; once multiplied by $1/l_m$, these give only subleading
corrections to $g(\theta)$. We thus find to leading order
\beq\label{eq:g_asy}
g\t = g(0)-\frac{1}{l_m}\hs(\rho\N l_m^2 \sin\theta) \simeq a-b\sin\theta
\eeq
where, using Eq.~\eqref{eq:rho_scaling} and the leading-order relation
$\rho\N=\rho/3$
\bea
a\equiv\gnot&=&\frac{\ln^2l_m}{2w^2 l_m}+\order\left(\frac{\ln
    l_m}{l_m}\right)\label{eq:A0_sol}\\
b&=&\frac{8}{\pi}l_m\rho\N=\frac{8}{3\pi}\frac{\ln^2l_m}{2c_1w^2}+
\order(\ln l_m)\label{eq:A1_sol}
\eea
The dominance of the long rods in the nematic phase thus results in a
very simple form for $\gt$, with $a$ vanishing and $b$ slowly
diverging in the limit $l_m\rightarrow\infty$.

Having obtained the desired predictions from our theory, we can now
also verify that the assumption of a large dimensionless density
$\rhoeff$ for the nematic phase is justified. From
Eq.~\eqref{eq:rho_scaling} and the fact that to leading order
$\rho\N=\rho/3$ one has $\rhoeff=\rho\N l_m^2\sim l_m\ln^2 l_m$ for a
log-normal parent distribution, and this indeed becomes arbitrarily
large as the cutoff $l_m$ increases. The self-consistency of the other
assumption which we made, \ie, the dominance of the long rods, is
verified in App.~\ref{app:long_rods_dom}.

\subsection{Validity of Onsager theory}\label{sec:ons_validity}

Before comparing our theoretical predictions with numerical results at
finite cutoff, we briefly assess the limit of validity of Onsager's
second virial approximation.  For a monodisperse system, an analysis
of the scaling of the second and third virial
coefficients~\cite{VroLek92} shows that in the nematic phase typical
rod angles $\theta$ with the nematic axis have to be $\gg D/L$, with
$D$ and $L$ the diameter and length of the rods, for the truncation
after the second virial contribution to be justified. In our
situation, the nematic phase is effectively monodisperse with
(unnormalized) rod length $L_0 l_m$, so the condition becomes $\theta
\gg D/(L_0l_m)$. We showed above that the typical angles scale, for
large cutoff $l_m$, as $\theta\sim 1/\rhoeff=1/(\rho\N l_m^2)$, so that
the second virial approximation breaks down when
\beq
\frac{D}{L_0l_m}\sim\frac{1}{\rho\N l_m^2}
\label{eq:ons_valid}
\eeq
Now $\rho\N$ scales as $(\ln^2 l_m)/l_m$, so this becomes
$\ln^2l_m\sim L_0/D$. This shows that fairly large values of the
aspect ratio $L_0/D$ of the ``reference rods'' are necessary in order
for the theory to be valid for large cutoffs. For instance, one would
need $L_0/D\simeq 50$ for a cutoff $l_m=1000$. The longest rods are
then very thin indeed, with $L_0 l_m/D \simeq 50,000$.

A more physically intuitive interpretation of the above condition is
that it corresponds to the requirement of having a rod volume fraction
$\phi\ll 1$. For monodisperse rods, we have (see
App.~\ref{app:high_dens_onsager}) that at large dimensionless density
$\rho$, $\theta\sim\rho^{-1}\sim(L^2DN/V)^{-1}$. The limit of validity
of the Onsager theory is therefore given by $D/L\sim\theta\sim
V/(L^2DN)$ or $1\sim (LD^2N)/V\sim\phi$. The same is true for our
calculation above: the volume fraction of the nematic phase is
$\phi\N=(D/L_0)\rho_1\N \simeq (D/L_0)l_m\rho\N$. The
condition~\eqref{eq:ons_valid} thus again becomes $\phi\N \sim 1$, and
we need $\phi\N\ll 1$ for the second virial theory to be valid.

\subsection{Comparison with numerical results}
\label{sec:num_support}

We now compare the theoretical predictions obtained above for the
limit $l_m\to\infty$ with numerical calculations for finite but large
cutoff. Our numerical results will be able to confirm only the leading
terms of the scaling solution, since sub-leading corrections (\eg\ to the
result~\eqref{eq:g_asy} for $g(\theta)$) can arise from the regime
of very small angles $\theta\sim 1/\rhoeff=1/(\rho\N l_m^2)$ which we
cannot resolve numerically.

\begin{figure}[htb]
\begin{center}
\begin{picture}(0,0)%
\includegraphics{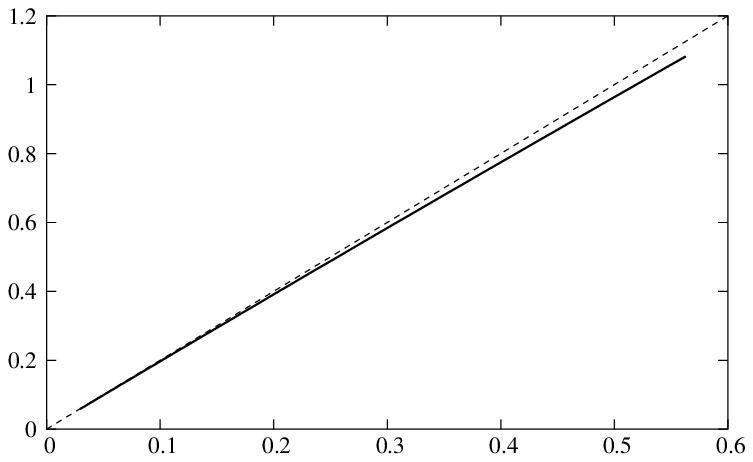}%
\end{picture}%
\setlength{\unitlength}{2447sp}%
\begingroup\makeatletter\ifx\SetFigFont\undefined
\def\x#1#2#3#4#5#6#7\relax{\def\x{#1#2#3#4#5#6}}%
\expandafter\x\fmtname xxxxxx\relax \def\y{splain}%
\ifx\x\y   
\gdef\SetFigFont#1#2#3{%
  \ifnum #1<17\tiny\else \ifnum #1<20\small\else
  \ifnum #1<24\normalsize\else \ifnum #1<29\large\else
  \ifnum #1<34\Large\else \ifnum #1<41\LARGE\else
     \huge\fi\fi\fi\fi\fi\fi
  \csname #3\endcsname}%
\else
\gdef\SetFigFont#1#2#3{\begingroup
  \count@#1\relax \ifnum 25<\count@\count@25\fi
  \def\x{\endgroup\@setsize\SetFigFont{#2pt}}%
  \expandafter\x
    \csname \romannumeral\the\count@ pt\expandafter\endcsname
    \csname @\romannumeral\the\count@ pt\endcsname
  \csname #3\endcsname}%
\fi
\fi\endgroup
\begin{picture}(5850,3788)(1276,-4235)
\put(1276,-1936){\makebox(0,0)[lb]{\smash{\SetFigFont{7}{8.4}{rm}{\color[rgb]{0,0,0}$\gnot$}%
}}}
\put(4351,-4186){\makebox(0,0)[b]{\smash{\SetFigFont{7}{8.4}{rm}{\color[rgb]{0,0,0}$\rho$}%
}}}
\end{picture}
\vspace*{0.3cm}
\caption{Parametric plot of $\gnot$ against $\rho$, for a log-normal
distribution with $\sigma=0.5$ and a 
  range of cutoffs between 50 and 3000. At small $\rho$ (large $l_m$)
  the numerical results (solid) are in very good agreement with the
  theoretically predicted asymptotic relation $\gnot=c_1\rho$ (dashed).}
\label{fig:rho_g0}
\efig
We begin by checking the predicted relations between the cloud point
density $\rho$ and the other quantities we have analysed
theoretically, namely the parameters $a$ and $b$ specifying the
leading behavior~\eqref{eq:g_asy} of $g(\theta)$, and the nematic
shadow density $\rho\N$. In Fig.~\ref{fig:rho_g0} we plot $\gnot\equiv
a$ against $\rho$ for a range of cutoffs $l_m$ between 50 and
3000. At large cutoff, \ie, at small $\rho$, the numerical results are
clearly seen to approach the theoretical prediction $g(0)=c_1\rho$,
while for smaller cutoffs deviations from the asymptotic theory appear
as expected.

For the parameter $b$, our theory predicted $b = (8/3\pi)l_m\rho$ in
the limit of large $l_m$.
\bfig
\begin{picture}(0,0)%
\includegraphics{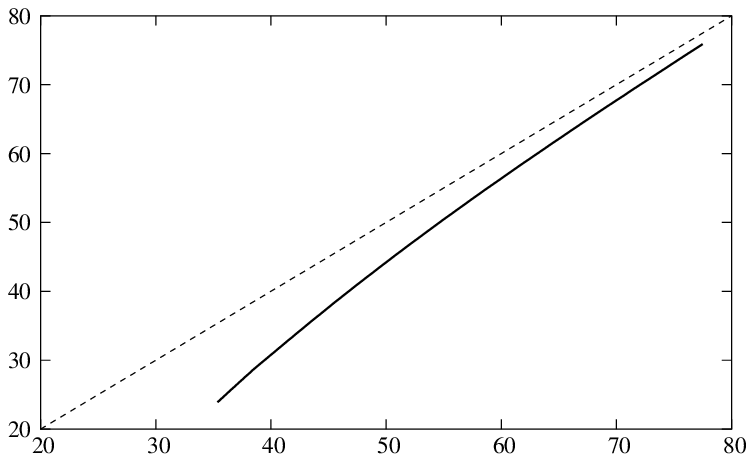}%
\end{picture}%
\setlength{\unitlength}{2447sp}%
\begingroup\makeatletter\ifx\SetFigFont\undefined
\def\x#1#2#3#4#5#6#7\relax{\def\x{#1#2#3#4#5#6}}%
\expandafter\x\fmtname xxxxxx\relax \def\y{splain}%
\ifx\x\y   
\gdef\SetFigFont#1#2#3{%
  \ifnum #1<17\tiny\else \ifnum #1<20\small\else
  \ifnum #1<24\normalsize\else \ifnum #1<29\large\else
  \ifnum #1<34\Large\else \ifnum #1<41\LARGE\else
     \huge\fi\fi\fi\fi\fi\fi
  \csname #3\endcsname}%
\else
\gdef\SetFigFont#1#2#3{\begingroup
  \count@#1\relax \ifnum 25<\count@\count@25\fi
  \def\x{\endgroup\@setsize\SetFigFont{#2pt}}%
  \expandafter\x
    \csname \romannumeral\the\count@ pt\expandafter\endcsname
    \csname @\romannumeral\the\count@ pt\endcsname
  \csname #3\endcsname}%
\fi
\fi\endgroup
\begin{picture}(5897,3797)(1201,-4244)
\put(1201,-1936){\makebox(0,0)[lb]{\smash{\SetFigFont{7}{8.4}{rm}{\color[rgb]{0,0,0}$b$}%
}}}
\put(4313,-4186){\makebox(0,0)[b]{\smash{\SetFigFont{7}{8.4}{rm}{\color[rgb]{0,0,0}$(8/3\pi)l_m\rho$}%
}}}
\end{picture}
\vspace*{0.3cm}
\caption{Parametric plot of $b$ against the theoretically predicted value
  $(8/3\pi)l_m\rho$ for a log-normal distribution with $\sigma=0.5$
and cutoff between 50 and 3000. The convergence of the numerical
results (solid) to the theoretical prediction (dashed) for large
cutoff, \ie, large $l_m\rho$, is clear.}
\label{fig:A1_rho}
\efig
In Fig.~\ref{fig:A1_rho} we plot the numerically obtained $b$ against
this theoretical prediction, for a range of cutoffs between 50 and
3000. At large $l_m\rho$ the convergence to the theoretical solution
is clear, while at finite cutoff (small $l_m\rho$) the value of $b$
lies rather below the theoretical expectation. This is not surprising,
as $b$ is expected to diverge with $l_m$ like $\ln^2 l_m$ and
therefore rather slowly; at finite cutoff, then, correction terms will
be rather important.

Based on the fact that the nematic pressure obeys the simple relation
$\Pi\N=3\rho\N$ in the large cutoff limit, our theory also predicts
that $\rho\N$ should be related to $\rho$ by
$\rho\N=[\rho+(c_1/2)\rho^2]/3=(\rho+\rho^2)/3$. A plot
of $\rho\N$ against $\rho$ for the same range of cutoffs as above
(Fig.~\ref{fig:press_eq_asy}) clearly shows that this relation is
satisfied in the limit of large $l_m$, \ie, small $\rho$.
\bfig
\begin{picture}(0,0)%
\includegraphics{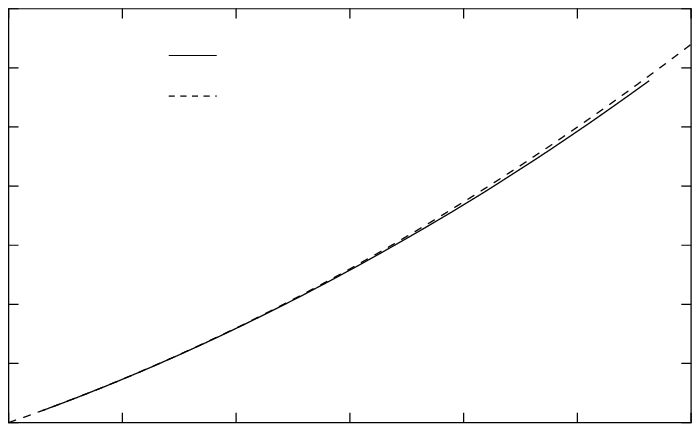}%
\end{picture}%
\setlength{\unitlength}{2447sp}%
\begingroup\makeatletter\ifx\SetFigFont\undefined
\def\x#1#2#3#4#5#6#7\relax{\def\x{#1#2#3#4#5#6}}%
\expandafter\x\fmtname xxxxxx\relax \def\y{splain}%
\ifx\x\y   
\gdef\SetFigFont#1#2#3{%
  \ifnum #1<17\tiny\else \ifnum #1<20\small\else
  \ifnum #1<24\normalsize\else \ifnum #1<29\large\else
  \ifnum #1<34\Large\else \ifnum #1<41\LARGE\else
     \huge\fi\fi\fi\fi\fi\fi
  \csname #3\endcsname}%
\else
\gdef\SetFigFont#1#2#3{\begingroup
  \count@#1\relax \ifnum 25<\count@\count@25\fi
  \def\x{\endgroup\@setsize\SetFigFont{#2pt}}%
  \expandafter\x
    \csname \romannumeral\the\count@ pt\expandafter\endcsname
    \csname @\romannumeral\the\count@ pt\endcsname
  \csname #3\endcsname}%
\fi
\fi\endgroup
\begin{picture}(5370,3705)(733,-3250)
\put(3449,-3201){\makebox(0,0)[b]{\smash{\SetFigFont{7}{8.4}{rm}{\color[rgb]{0,0,0}$\rho$}%
}}}
\put(1973,-33){\makebox(0,0)[rb]{\smash{\SetFigFont{7}{8.4}{rm}{\color[rgb]{0,0,0}$\rho\N$}%
}}}
\put(1973,-336){\makebox(0,0)[rb]{\smash{\SetFigFont{7}{8.4}{rm}$\frac{1}{3}\left(\rho+\rho^2\right)$}}}
\put(733,-2875){\makebox(0,0)[rb]{\smash{\SetFigFont{7}{8.4}{rm}{\color[rgb]{0,0,0}0}%
}}}
\put(733,-2417){\makebox(0,0)[rb]{\smash{\SetFigFont{7}{8.4}{rm}{\color[rgb]{0,0,0}0.05}%
}}}
\put(733,-1960){\makebox(0,0)[rb]{\smash{\SetFigFont{7}{8.4}{rm}{\color[rgb]{0,0,0}0.1}%
}}}
\put(733,-1502){\makebox(0,0)[rb]{\smash{\SetFigFont{7}{8.4}{rm}{\color[rgb]{0,0,0}0.15}%
}}}
\put(733,-1044){\makebox(0,0)[rb]{\smash{\SetFigFont{7}{8.4}{rm}{\color[rgb]{0,0,0}0.2}%
}}}
\put(733,-586){\makebox(0,0)[rb]{\smash{\SetFigFont{7}{8.4}{rm}{\color[rgb]{0,0,0}0.25}%
}}}
\put(733,-129){\makebox(0,0)[rb]{\smash{\SetFigFont{7}{8.4}{rm}{\color[rgb]{0,0,0}0.3}%
}}}
\put(733,329){\makebox(0,0)[rb]{\smash{\SetFigFont{7}{8.4}{rm}{\color[rgb]{0,0,0}0.35}%
}}}
\put(807,-2999){\makebox(0,0)[b]{\smash{\SetFigFont{7}{8.4}{rm}{\color[rgb]{0,0,0}0}%
}}}
\put(1688,-2999){\makebox(0,0)[b]{\smash{\SetFigFont{7}{8.4}{rm}{\color[rgb]{0,0,0}0.1}%
}}}
\put(2568,-2999){\makebox(0,0)[b]{\smash{\SetFigFont{7}{8.4}{rm}{\color[rgb]{0,0,0}0.2}%
}}}
\put(4330,-2999){\makebox(0,0)[b]{\smash{\SetFigFont{7}{8.4}{rm}{\color[rgb]{0,0,0}0.4}%
}}}
\put(5210,-2999){\makebox(0,0)[b]{\smash{\SetFigFont{7}{8.4}{rm}{\color[rgb]{0,0,0}0.5}%
}}}
\put(6091,-2999){\makebox(0,0)[b]{\smash{\SetFigFont{7}{8.4}{rm}{\color[rgb]{0,0,0}0.6}%
}}}
\put(3449,-2999){\makebox(0,0)[b]{\smash{\SetFigFont{7}{8.4}{rm}{\color[rgb]{0,0,0}0.3}%
}}}
\end{picture}
\vspace*{0.3cm}
\caption{Parametric plot of the nematic
shadow density $\rho\N$ against $\rho$, for the same parent
distribution and cutoff range as in Figs.~\protect\ref{fig:rho_g0}
and~\protect\ref{fig:A1_rho}. Very good agreement is observed between
the numerical results (solid) and the theoretically predicted relation
(dashed).}
\label{fig:press_eq_asy}
\efig
In fact, deviations from the predicted scaling are rather small
already at modest $l_m$ (here in the range $50\leq l_m\leq
3000$). This shows that the dominance of the long rods, demonstrated
also by the fact that the average length of the nematic phase is very
close to $l_m$ (Fig.~\ref{fig:length_50_100}), appears quite
early on. Even at $l_m=50$, not only is the nematic phase composed almost
entirely of rods of order $l_m$, but the rods are also sufficiently
strongly ordered to make our scaling solution a good approximation.

Having confirmed the relations between $a, b,\rho$ and $\rho\N$, the
last prediction to verify is the variation of one of these quantities
with $l_m$. We choose $a\equiv\gnot$, for which our theory predicts
the leading asymptotic behavior $\gnot = (\ln^2l_m)/(2 w^2l_m)$.
\begin{figure}[htb]
\begin{center}
\begin{picture}(0,0)%
\includegraphics{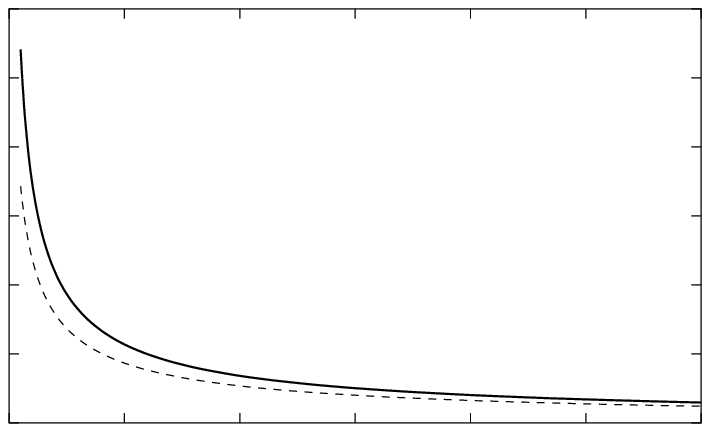}%
\end{picture}%
\setlength{\unitlength}{2447sp}%
\begingroup\makeatletter\ifx\SetFigFont\undefined
\def\x#1#2#3#4#5#6#7\relax{\def\x{#1#2#3#4#5#6}}%
\expandafter\x\fmtname xxxxxx\relax \def\y{splain}%
\ifx\x\y   
\gdef\SetFigFont#1#2#3{%
  \ifnum #1<17\tiny\else \ifnum #1<20\small\else
  \ifnum #1<24\normalsize\else \ifnum #1<29\large\else
  \ifnum #1<34\Large\else \ifnum #1<41\LARGE\else
     \huge\fi\fi\fi\fi\fi\fi
  \csname #3\endcsname}%
\else
\gdef\SetFigFont#1#2#3{\begingroup
  \count@#1\relax \ifnum 25<\count@\count@25\fi
  \def\x{\endgroup\@setsize\SetFigFont{#2pt}}%
  \expandafter\x
    \csname \romannumeral\the\count@ pt\expandafter\endcsname
    \csname @\romannumeral\the\count@ pt\endcsname
  \csname #3\endcsname}%
\fi
\fi\endgroup
\begin{picture}(5962,3795)(151,-3340)
\put(3376,-3286){\makebox(0,0)[lb]{\smash{\SetFigFont{7}{8.4}{rm}{\color[rgb]{0,0,0}$l_m$}%
}}}
\put(151,-1036){\makebox(0,0)[lb]{\smash{\SetFigFont{7}{8.4}{rm}{\color[rgb]{0,0,0}$\gnot$}%
}}}
\put(659,-2875){\makebox(0,0)[rb]{\smash{\SetFigFont{7}{8.4}{rm}{\color[rgb]{0,0,0}0}%
}}}
\put(659,-2341){\makebox(0,0)[rb]{\smash{\SetFigFont{7}{8.4}{rm}{\color[rgb]{0,0,0}0.2}%
}}}
\put(659,-1807){\makebox(0,0)[rb]{\smash{\SetFigFont{7}{8.4}{rm}{\color[rgb]{0,0,0}0.4}%
}}}
\put(659,-1273){\makebox(0,0)[rb]{\smash{\SetFigFont{7}{8.4}{rm}{\color[rgb]{0,0,0}0.6}%
}}}
\put(659,-739){\makebox(0,0)[rb]{\smash{\SetFigFont{7}{8.4}{rm}{\color[rgb]{0,0,0}0.8}%
}}}
\put(659,-205){\makebox(0,0)[rb]{\smash{\SetFigFont{7}{8.4}{rm}{\color[rgb]{0,0,0}1}%
}}}
\put(659,329){\makebox(0,0)[rb]{\smash{\SetFigFont{7}{8.4}{rm}{\color[rgb]{0,0,0}1.2}%
}}}
\put(733,-2999){\makebox(0,0)[b]{\smash{\SetFigFont{7}{8.4}{rm}{\color[rgb]{0,0,0}0}%
}}}
\put(1626,-2999){\makebox(0,0)[b]{\smash{\SetFigFont{7}{8.4}{rm}{\color[rgb]{0,0,0}500}%
}}}
\put(2519,-2999){\makebox(0,0)[b]{\smash{\SetFigFont{7}{8.4}{rm}{\color[rgb]{0,0,0}1000}%
}}}
\put(3412,-2999){\makebox(0,0)[b]{\smash{\SetFigFont{7}{8.4}{rm}{\color[rgb]{0,0,0}1500}%
}}}
\put(4305,-2999){\makebox(0,0)[b]{\smash{\SetFigFont{7}{8.4}{rm}{\color[rgb]{0,0,0}2000}%
}}}
\put(5198,-2999){\makebox(0,0)[b]{\smash{\SetFigFont{7}{8.4}{rm}{\color[rgb]{0,0,0}2500}%
}}}
\put(6091,-2999){\makebox(0,0)[b]{\smash{\SetFigFont{7}{8.4}{rm}{\color[rgb]{0,0,0}3000}%
}}}
\end{picture}
\vspace*{0.3cm}
\caption{Plot of numerically calculated values of $\gnot$ against
$l_m$ (solid) together with the leading asymptotic behavior
$g(0)=(\ln^2l_m)/(2w^2l_m)$ predicted theoretically (dashed), for a
log-normal distribution with $\sigma=0.5$.  The agreement is
satisfactory, though corrections to the asymptotic theory
are clearly still important in the range of $l_m$ shown.
}
\label{fig:g0_scaling_both}
\efig
In Fig.~\ref{fig:g0_scaling_both}, we plot the numerically calculated
values of $\gnot$ versus $l_m$ and compare with the theoretical
prediction. The overall shape of the $l_m$-dependence is well captured
by the theory, although subleading corrections, which from
Eq.~\eqref{eq:gnot_scaling} are of relative order $\order(1/\ln l_m)$,
are clearly not yet negligible in the range of $l_m$ considered.
%
%
In summary, then, all numerical results are consistent with the
theoretical predictions derived above.

\section{The Schulz distribution}\label{sec:schulz}


Having observed the rather surprising effects caused by the long rods
at the onset of isotropic-nematic coexistence in systems with
fat-tailed length distributions, an obvious question is whether
similar phenomena are possible even for more strongly decaying length
distributions. We therefore now analyse, using the same numerical and
theoretical methods as above, the case of a Schulz distribution of
lengths. For this distribution our previous studies of the
Zwanzig~\cite{ClaCueSeaSolSpe00} and $\Ptwo$ Onsager
models~\cite{SpeSol_p2} did not show any signs of the phase behavior
being driven by the long rods. However, comparing our above results
for the log-normal case with those obtained for the $\Ptwo$ Onsager
model~\cite{SpeSol_p2_fat}, it is clear that in the full Onsager
theory the effect of the long rods is much more pronounced than in the
approximate models. Long rod effects might therefore also appear, in
the full Onsager theory, for the more strongly decaying Schulz
distribution, but would then be expected to be weaker than for the
log-normal case.

The Schulz length distribution can be written as 
\beq\label{eq:schulz_parent}
\normparent(l)=\frac{(z+1)^{z+1}}{\Gamma(z+1)}\, l^z\exp[-(z+1)l]
\eeq
where we have again imposed an average length of 1 (for infinite
cutoff $l_m$). The polydispersity $\sigma$, defined as before as the
relative standard deviation of the distribution, is related to $z$ by
\begin{equation}\label{eq:sigma}
\sigma^2=\frac{\langle l^2\rangle - \langle l \rangle^2}
{\langle l\rangle^2}
=\frac{1}{z+1}
\end{equation}
From Eqs.~\eqref{eq:fat_condition} and~\eqref{eq:schulz_parent} it
is clear that if $\gnot\geq z+1$ the nematic density distribution is
again exponentially diverging for large $l$. Assuming initially that
this is not the case, however, one can solve numerically
Eqs.~\eqref{eq:g_eq} and~\eqref{eq:Pi_eq} for the onset of
isotropic-nematic phase coexistence. The results for $g(0)$ are shown in
Fig.~\ref{fig:g0_schulz}. For large $z$, \ie, small polydispersity
$\sigma=(1+z)^{-1/2}$, one has $g(0)<z+1$ (dashed line). In this
regime the nematic density distribution decays exponentially for large
$l$, and the results are essentially independent of the cutoff $l_m$,
which could in fact be taken to infinity. For smaller $z$, on the other hand,
$\gnot>z+1$, implying that the nematic density distribution is
exponentially increasing and a finite cutoff $l_m$ is necessary.
\begin{figure}[htb]
\begin{center}
\begin{picture}(0,0)%
\includegraphics{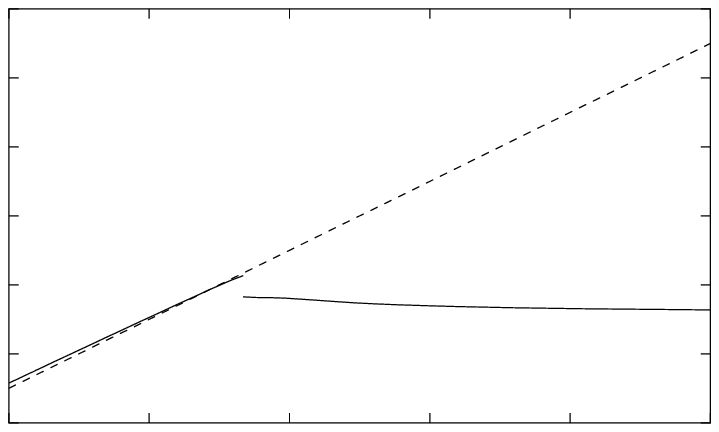}%
\end{picture}%
\setlength{\unitlength}{2447sp}%
\begingroup\makeatletter\ifx\SetFigFont\undefined
\def\x#1#2#3#4#5#6#7\relax{\def\x{#1#2#3#4#5#6}}%
\expandafter\x\fmtname xxxxxx\relax \def\y{splain}%
\ifx\x\y   
\gdef\SetFigFont#1#2#3{%
  \ifnum #1<17\tiny\else \ifnum #1<20\small\else
  \ifnum #1<24\normalsize\else \ifnum #1<29\large\else
  \ifnum #1<34\Large\else \ifnum #1<41\LARGE\else
     \huge\fi\fi\fi\fi\fi\fi
  \csname #3\endcsname}%
\else
\gdef\SetFigFont#1#2#3{\begingroup
  \count@#1\relax \ifnum 25<\count@\count@25\fi
  \def\x{\endgroup\@setsize\SetFigFont{#2pt}}%
  \expandafter\x
    \csname \romannumeral\the\count@ pt\expandafter\endcsname
    \csname @\romannumeral\the\count@ pt\endcsname
  \csname #3\endcsname}%
\fi
\fi\endgroup
\begin{picture}(5877,3665)(226,-3210)
\put(226,-961){\makebox(0,0)[lb]{\smash{\SetFigFont{7}{8.4}{rm}{\color[rgb]{0,0,0}$g(0)$}%
}}}
\put(3375,-3161){\makebox(0,0)[b]{\smash{\SetFigFont{7}{8.4}{rm}{\color[rgb]{0,0,0}$z$}%
}}}
\put(585,-2875){\makebox(0,0)[rb]{\smash{\SetFigFont{7}{8.4}{rm}{\color[rgb]{0,0,0}0}%
}}}
\put(585,-2341){\makebox(0,0)[rb]{\smash{\SetFigFont{7}{8.4}{rm}{\color[rgb]{0,0,0}2}%
}}}
\put(585,-1807){\makebox(0,0)[rb]{\smash{\SetFigFont{7}{8.4}{rm}{\color[rgb]{0,0,0}4}%
}}}
\put(585,-1273){\makebox(0,0)[rb]{\smash{\SetFigFont{7}{8.4}{rm}{\color[rgb]{0,0,0}6}%
}}}
\put(585,-739){\makebox(0,0)[rb]{\smash{\SetFigFont{7}{8.4}{rm}{\color[rgb]{0,0,0}8}%
}}}
\put(585,-205){\makebox(0,0)[rb]{\smash{\SetFigFont{7}{8.4}{rm}{\color[rgb]{0,0,0}10}%
}}}
\put(585,329){\makebox(0,0)[rb]{\smash{\SetFigFont{7}{8.4}{rm}{\color[rgb]{0,0,0}12}%
}}}
\put(659,-2999){\makebox(0,0)[b]{\smash{\SetFigFont{7}{8.4}{rm}{\color[rgb]{0,0,0}0}%
}}}
\put(1745,-2999){\makebox(0,0)[b]{\smash{\SetFigFont{7}{8.4}{rm}{\color[rgb]{0,0,0}2}%
}}}
\put(5005,-2999){\makebox(0,0)[b]{\smash{\SetFigFont{7}{8.4}{rm}{\color[rgb]{0,0,0}8}%
}}}
\put(6091,-2999){\makebox(0,0)[b]{\smash{\SetFigFont{7}{8.4}{rm}{\color[rgb]{0,0,0}10}%
}}}
\put(3918,-2999){\makebox(0,0)[b]{\smash{\SetFigFont{7}{8.4}{rm}{\color[rgb]{0,0,0}6}%
}}}
\put(2832,-2999){\makebox(0,0)[b]{\smash{\SetFigFont{7}{8.4}{rm}{\color[rgb]{0,0,0}4}%
}}}
\end{picture}
\vspace*{0.3cm}
\caption{The parameter $\gnot$ of the nematic shadow phase, plotted
against $z$ for a Schulz distribution with cutoff $l_m=100$. See text
for discussion.
}
\label{fig:g0_schulz}
\efig
In this regime, the situation resembles the case of the fat-tailed
length distributions discussed earlier, where for large enough cutoff
and polydispersity the less than exponentially decaying length
distribution was not able to balance the divergence of the factor
$\exp[{l\gnot}]$ in Eq.~\eqref{eq:fat_condition}. With a Schulz
distribution, the only difference is that now the comparison is
between two exponential terms $\exp[-(z+1)l]$ and
$\exp[l\gnot]$. Given this analogy, it is not surprising that the
cloud and shadow curves (Fig.~\ref{fig:cloud_shad_50_100_schulz}) show
behavior qualitatively similar to that found for the log-normal
distribution: a kink in the cloud curve and a discontinuity in the
shadow curve again indicate the presence of a three-phase I-N-N
coexistence region in the phase diagram.
\begin{figure}[h!]
\begin{center}
\begin{picture}(0,0)%
\includegraphics{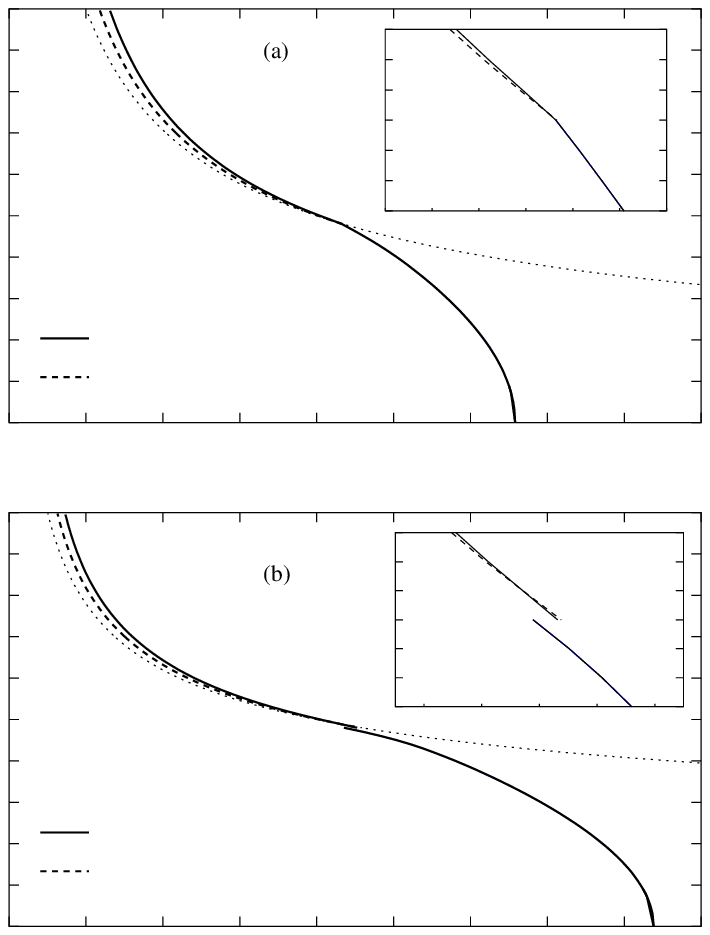}%
\end{picture}%
\setlength{\unitlength}{2447sp}%
\begingroup\makeatletter\ifx\SetFigFont\undefined
\def\x#1#2#3#4#5#6#7\relax{\def\x{#1#2#3#4#5#6}}%
\expandafter\x\fmtname xxxxxx\relax \def\y{splain}%
\ifx\x\y   
\gdef\SetFigFont#1#2#3{%
  \ifnum #1<17\tiny\else \ifnum #1<20\small\else
  \ifnum #1<24\normalsize\else \ifnum #1<29\large\else
  \ifnum #1<34\Large\else \ifnum #1<41\LARGE\else
     \huge\fi\fi\fi\fi\fi\fi
  \csname #3\endcsname}%
\else
\gdef\SetFigFont#1#2#3{\begingroup
  \count@#1\relax \ifnum 25<\count@\count@25\fi
  \def\x{\endgroup\@setsize\SetFigFont{#2pt}}%
  \expandafter\x
    \csname \romannumeral\the\count@ pt\expandafter\endcsname
    \csname @\romannumeral\the\count@ pt\endcsname
  \csname #3\endcsname}%
\fi
\fi\endgroup
\begin{picture}(5952,7690)(151,-7235)
\put(1426,-2236){\makebox(0,0)[lb]{\smash{\SetFigFont{7}{8.4}{rm}{\color[rgb]{0,0,0}$l_m=50$}%
}}}
\put(1426,-2536){\makebox(0,0)[lb]{\smash{\SetFigFont{7}{8.4}{rm}{\color[rgb]{0,0,0}$l_m=100$}%
}}}
\put(1426,-6061){\makebox(0,0)[lb]{\smash{\SetFigFont{7}{8.4}{rm}{\color[rgb]{0,0,0}$l_m=50$}%
}}}
\put(1426,-6361){\makebox(0,0)[lb]{\smash{\SetFigFont{7}{8.4}{rm}{\color[rgb]{0,0,0}$l_m=100$}%
}}}
\put(151,-1261){\makebox(0,0)[lb]{\smash{\SetFigFont{7}{8.4}{rm}{\color[rgb]{0,0,0}$\sigma$}%
}}}
\put(3301,-7186){\makebox(0,0)[lb]{\smash{\SetFigFont{7}{8.4}{rm}{\color[rgb]{0,0,0}$\rho\N$}%
}}}
\put(226,-5161){\makebox(0,0)[lb]{\smash{\SetFigFont{7}{8.4}{rm}{\color[rgb]{0,0,0}$\sigma$}%
}}}
\put(3412,-3161){\makebox(0,0)[b]{\smash{\SetFigFont{7}{8.4}{rm}{\color[rgb]{0,0,0}$\rho$}%
}}}
\put(3666,-4805){\makebox(0,0)[rb]{\smash{\SetFigFont{7}{8.4}{rm}{\color[rgb]{0,0,0}0.46}%
}}}
\put(3666,-4357){\makebox(0,0)[rb]{\smash{\SetFigFont{7}{8.4}{rm}{\color[rgb]{0,0,0}0.48}%
}}}
\put(3666,-3908){\makebox(0,0)[rb]{\smash{\SetFigFont{7}{8.4}{rm}{\color[rgb]{0,0,0}0.5}%
}}}
\put(4392,-5161){\makebox(0,0)[b]{\smash{\SetFigFont{7}{8.4}{rm}{\color[rgb]{0,0,0}2.0}%
}}}
\put(5285,-5161){\makebox(0,0)[b]{\smash{\SetFigFont{7}{8.4}{rm}{\color[rgb]{0,0,0}2.4}%
}}}
\put(4008,-1357){\makebox(0,0)[b]{\smash{\SetFigFont{7}{8.4}{rm}{\color[rgb]{0,0,0}1.9}%
}}}
\put(4734,-1357){\makebox(0,0)[b]{\smash{\SetFigFont{7}{8.4}{rm}{\color[rgb]{0,0,0}2.1}%
}}}
\put(5460,-1357){\makebox(0,0)[b]{\smash{\SetFigFont{7}{8.4}{rm}{\color[rgb]{0,0,0}2.3}%
}}}
\put(3588,-471){\makebox(0,0)[rb]{\smash{\SetFigFont{7}{8.4}{rm}{\color[rgb]{0,0,0}0.48}%
}}}
\put(3588,-940){\makebox(0,0)[rb]{\smash{\SetFigFont{7}{8.4}{rm}{\color[rgb]{0,0,0}0.46}%
}}}
\put(3588, -3){\makebox(0,0)[rb]{\smash{\SetFigFont{7}{8.4}{rm}{\color[rgb]{0,0,0}0.5}%
}}}
\put(659,-2875){\makebox(0,0)[rb]{\smash{\SetFigFont{7}{8.4}{rm}{\color[rgb]{0,0,0}0}%
}}}
\put(659,-2555){\makebox(0,0)[rb]{\smash{\SetFigFont{7}{8.4}{rm}{\color[rgb]{0,0,0}0.1}%
}}}
\put(659,-2234){\makebox(0,0)[rb]{\smash{\SetFigFont{7}{8.4}{rm}{\color[rgb]{0,0,0}0.2}%
}}}
\put(659,-1914){\makebox(0,0)[rb]{\smash{\SetFigFont{7}{8.4}{rm}{\color[rgb]{0,0,0}0.3}%
}}}
\put(659,-1593){\makebox(0,0)[rb]{\smash{\SetFigFont{7}{8.4}{rm}{\color[rgb]{0,0,0}0.4}%
}}}
\put(659,-1273){\makebox(0,0)[rb]{\smash{\SetFigFont{7}{8.4}{rm}{\color[rgb]{0,0,0}0.5}%
}}}
\put(659,-953){\makebox(0,0)[rb]{\smash{\SetFigFont{7}{8.4}{rm}{\color[rgb]{0,0,0}0.6}%
}}}
\put(659,-632){\makebox(0,0)[rb]{\smash{\SetFigFont{7}{8.4}{rm}{\color[rgb]{0,0,0}0.7}%
}}}
\put(659,-312){\makebox(0,0)[rb]{\smash{\SetFigFont{7}{8.4}{rm}{\color[rgb]{0,0,0}0.8}%
}}}
\put(659,  9){\makebox(0,0)[rb]{\smash{\SetFigFont{7}{8.4}{rm}{\color[rgb]{0,0,0}0.9}%
}}}
\put(659,329){\makebox(0,0)[rb]{\smash{\SetFigFont{7}{8.4}{rm}{\color[rgb]{0,0,0}1}%
}}}
\put(733,-2999){\makebox(0,0)[b]{\smash{\SetFigFont{7}{8.4}{rm}{\color[rgb]{0,0,0}0}%
}}}
\put(1328,-2999){\makebox(0,0)[b]{\smash{\SetFigFont{7}{8.4}{rm}{\color[rgb]{0,0,0}0.5}%
}}}
\put(1924,-2999){\makebox(0,0)[b]{\smash{\SetFigFont{7}{8.4}{rm}{\color[rgb]{0,0,0}1}%
}}}
\put(2519,-2999){\makebox(0,0)[b]{\smash{\SetFigFont{7}{8.4}{rm}{\color[rgb]{0,0,0}1.5}%
}}}
\put(4305,-2999){\makebox(0,0)[b]{\smash{\SetFigFont{7}{8.4}{rm}{\color[rgb]{0,0,0}3}%
}}}
\put(4900,-2999){\makebox(0,0)[b]{\smash{\SetFigFont{7}{8.4}{rm}{\color[rgb]{0,0,0}3.5}%
}}}
\put(5496,-2999){\makebox(0,0)[b]{\smash{\SetFigFont{7}{8.4}{rm}{\color[rgb]{0,0,0}4}%
}}}
\put(6091,-2999){\makebox(0,0)[b]{\smash{\SetFigFont{7}{8.4}{rm}{\color[rgb]{0,0,0}4.5}%
}}}
\put(659,-6775){\makebox(0,0)[rb]{\smash{\SetFigFont{7}{8.4}{rm}{\color[rgb]{0,0,0}0}%
}}}
\put(659,-6455){\makebox(0,0)[rb]{\smash{\SetFigFont{7}{8.4}{rm}{\color[rgb]{0,0,0}0.1}%
}}}
\put(659,-6134){\makebox(0,0)[rb]{\smash{\SetFigFont{7}{8.4}{rm}{\color[rgb]{0,0,0}0.2}%
}}}
\put(659,-5814){\makebox(0,0)[rb]{\smash{\SetFigFont{7}{8.4}{rm}{\color[rgb]{0,0,0}0.3}%
}}}
\put(659,-5493){\makebox(0,0)[rb]{\smash{\SetFigFont{7}{8.4}{rm}{\color[rgb]{0,0,0}0.4}%
}}}
\put(659,-5173){\makebox(0,0)[rb]{\smash{\SetFigFont{7}{8.4}{rm}{\color[rgb]{0,0,0}0.5}%
}}}
\put(659,-4853){\makebox(0,0)[rb]{\smash{\SetFigFont{7}{8.4}{rm}{\color[rgb]{0,0,0}0.6}%
}}}
\put(659,-4532){\makebox(0,0)[rb]{\smash{\SetFigFont{7}{8.4}{rm}{\color[rgb]{0,0,0}0.7}%
}}}
\put(659,-4212){\makebox(0,0)[rb]{\smash{\SetFigFont{7}{8.4}{rm}{\color[rgb]{0,0,0}0.8}%
}}}
\put(659,-3891){\makebox(0,0)[rb]{\smash{\SetFigFont{7}{8.4}{rm}{\color[rgb]{0,0,0}0.9}%
}}}
\put(659,-3571){\makebox(0,0)[rb]{\smash{\SetFigFont{7}{8.4}{rm}{\color[rgb]{0,0,0}1}%
}}}
\put(733,-6899){\makebox(0,0)[b]{\smash{\SetFigFont{7}{8.4}{rm}{\color[rgb]{0,0,0}0}%
}}}
\put(1328,-6899){\makebox(0,0)[b]{\smash{\SetFigFont{7}{8.4}{rm}{\color[rgb]{0,0,0}0.5}%
}}}
\put(1924,-6899){\makebox(0,0)[b]{\smash{\SetFigFont{7}{8.4}{rm}{\color[rgb]{0,0,0}1}%
}}}
\put(2519,-6899){\makebox(0,0)[b]{\smash{\SetFigFont{7}{8.4}{rm}{\color[rgb]{0,0,0}1.5}%
}}}
\put(3114,-6899){\makebox(0,0)[b]{\smash{\SetFigFont{7}{8.4}{rm}{\color[rgb]{0,0,0}2}%
}}}
\put(3710,-6899){\makebox(0,0)[b]{\smash{\SetFigFont{7}{8.4}{rm}{\color[rgb]{0,0,0}2.5}%
}}}
\put(4305,-6899){\makebox(0,0)[b]{\smash{\SetFigFont{7}{8.4}{rm}{\color[rgb]{0,0,0}3}%
}}}
\put(4900,-6899){\makebox(0,0)[b]{\smash{\SetFigFont{7}{8.4}{rm}{\color[rgb]{0,0,0}3.5}%
}}}
\put(5496,-6899){\makebox(0,0)[b]{\smash{\SetFigFont{7}{8.4}{rm}{\color[rgb]{0,0,0}4}%
}}}
\put(6091,-6899){\makebox(0,0)[b]{\smash{\SetFigFont{7}{8.4}{rm}{\color[rgb]{0,0,0}4.5}%
}}}
\put(3114,-2999){\makebox(0,0)[b]{\smash{\SetFigFont{7}{8.4}{rm}{\color[rgb]{0,0,0}2}%
}}}
\put(3710,-2999){\makebox(0,0)[b]{\smash{\SetFigFont{7}{8.4}{rm}{\color[rgb]{0,0,0}2.5}%
}}}
\end{picture}
\vspace*{0.3cm}
\caption{(a) Cloud curves for Schulz distributions with cutoff
$l_m=50$ (solid) and $l_m=100$ (dashed), plotted as cloud point
  density $\rho$ versus polydispersity $\sigma$ on the $y$-axis. The
  system is again dominated by the long rods for large $\sigma$, but
  now the dependence on the cutoff is much less pronounced than for a
  log-normal length distribution
  (Fig.~\protect\ref{fig:cloud_shad_50_100}). The cloud curves also
  exhibit a kink, more clearly visible only on a magnified
  scale (inset). The dotted line represents the theoretically
  predicted limiting form $\rho=1/(c_1\sigma^2)$ of the cloud point
  curve above the kink (see text) in the limit $l_m\to\infty$. (b)
  Corresponding shadow curves.  The discontinuity corresponding to the
  kink in the cloud curves is much narrower than in the log-normal
  case (Fig.~\protect\ref{fig:cloud_shad_50_100}), and is visible
  clearly only in the inset. The dotted line again gives the
  theoretical prediction for the shadow curve above the discontinuity
  in the limit $l_m\to\infty$,
  $\rho\N=(1/\sigma^2+1/2\sigma^4)/(3c_1)$.}
\label{fig:cloud_shad_50_100_schulz}
\vspace*{-0.6cm}
\efig
Quantitatively, however, the kink in the cloud curve
(Fig.~\ref{fig:cloud_shad_50_100_schulz}(a)) is now much less
pronounced, and the cutoff-dependence of the cloud curve above the
kink is also
rather weaker. Similar comments apply to the shadow curve
(Fig.~\ref{fig:cloud_shad_50_100_schulz}(b)): the discontinuity is
still present but very small, with the nematic phases on the two
different branches having very similar densities
(Fig.~\ref{fig:cloud_shad_50_100_schulz}(b), inset).  As for the
log-normal distribution, for small polydispersities (\ie, below the
kink or discontinuity, respectively) the cloud and shadow curves are
essentially independent of the cutoff and connect smoothly with the
monodisperse limit at $\sigma=0$.

\begin{figure}[htb]
\begin{center}
\begin{picture}(0,0)%
\includegraphics{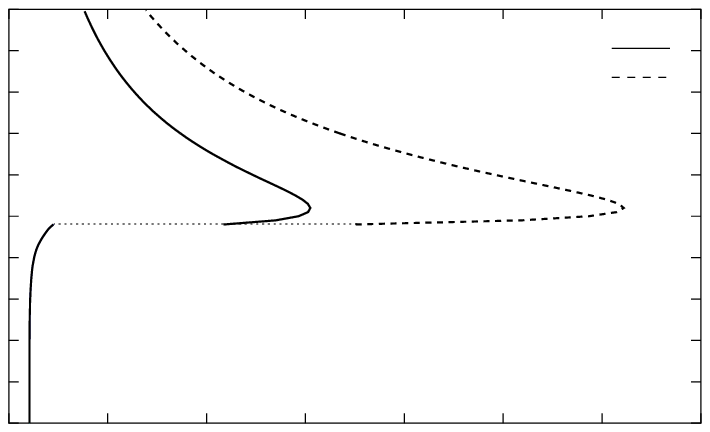}%
\end{picture}%
\setlength{\unitlength}{2447sp}%
\begingroup\makeatletter\ifx\SetFigFont\undefined
\def\x#1#2#3#4#5#6#7\relax{\def\x{#1#2#3#4#5#6}}%
\expandafter\x\fmtname xxxxxx\relax \def\y{splain}%
\ifx\x\y   
\gdef\SetFigFont#1#2#3{%
  \ifnum #1<17\tiny\else \ifnum #1<20\small\else
  \ifnum #1<24\normalsize\else \ifnum #1<29\large\else
  \ifnum #1<34\Large\else \ifnum #1<41\LARGE\else
     \huge\fi\fi\fi\fi\fi\fi
  \csname #3\endcsname}%
\else
\gdef\SetFigFont#1#2#3{\begingroup
  \count@#1\relax \ifnum 25<\count@\count@25\fi
  \def\x{\endgroup\@setsize\SetFigFont{#2pt}}%
  \expandafter\x
    \csname \romannumeral\the\count@ pt\expandafter\endcsname
    \csname @\romannumeral\the\count@ pt\endcsname
  \csname #3\endcsname}%
\fi
\fi\endgroup
\begin{picture}(6027,3724)(76,-3269)
\put(3376,-3211){\makebox(0,0)[lb]{\smash{\SetFigFont{7}{8.4}{rm}{\color[rgb]{0,0,0}$\rho_1\N$}%
}}}
\put( 76,-1111){\makebox(0,0)[lb]{\smash{\SetFigFont{7}{8.4}{rm}{\color[rgb]{0,0,0}$\sigma$}%
}}}
\put(4451,-211){\makebox(0,0)[lb]{\smash{\SetFigFont{7}{8.4}{rm}{\color[rgb]{0,0,0}$l_m=100$}%
}}}
\put(4451, 14){\makebox(0,0)[lb]{\smash{\SetFigFont{7}{8.4}{rm}{\color[rgb]{0,0,0}$l_m=50$}%
}}}
\put(659,-2875){\makebox(0,0)[rb]{\smash{\SetFigFont{7}{8.4}{rm}{\color[rgb]{0,0,0}0}%
}}}
\put(659,-2555){\makebox(0,0)[rb]{\smash{\SetFigFont{7}{8.4}{rm}{\color[rgb]{0,0,0}0.1}%
}}}
\put(659,-2234){\makebox(0,0)[rb]{\smash{\SetFigFont{7}{8.4}{rm}{\color[rgb]{0,0,0}0.2}%
}}}
\put(659,-1914){\makebox(0,0)[rb]{\smash{\SetFigFont{7}{8.4}{rm}{\color[rgb]{0,0,0}0.3}%
}}}
\put(659,-1593){\makebox(0,0)[rb]{\smash{\SetFigFont{7}{8.4}{rm}{\color[rgb]{0,0,0}0.4}%
}}}
\put(659,-1273){\makebox(0,0)[rb]{\smash{\SetFigFont{7}{8.4}{rm}{\color[rgb]{0,0,0}0.5}%
}}}
\put(659,-953){\makebox(0,0)[rb]{\smash{\SetFigFont{7}{8.4}{rm}{\color[rgb]{0,0,0}0.6}%
}}}
\put(659,-632){\makebox(0,0)[rb]{\smash{\SetFigFont{7}{8.4}{rm}{\color[rgb]{0,0,0}0.7}%
}}}
\put(659,-312){\makebox(0,0)[rb]{\smash{\SetFigFont{7}{8.4}{rm}{\color[rgb]{0,0,0}0.8}%
}}}
\put(659,  9){\makebox(0,0)[rb]{\smash{\SetFigFont{7}{8.4}{rm}{\color[rgb]{0,0,0}0.9}%
}}}
\put(659,329){\makebox(0,0)[rb]{\smash{\SetFigFont{7}{8.4}{rm}{\color[rgb]{0,0,0}1}%
}}}
\put(733,-2999){\makebox(0,0)[b]{\smash{\SetFigFont{7}{8.4}{rm}{\color[rgb]{0,0,0}0}%
}}}
\put(1498,-2999){\makebox(0,0)[b]{\smash{\SetFigFont{7}{8.4}{rm}{\color[rgb]{0,0,0}20}%
}}}
\put(2264,-2999){\makebox(0,0)[b]{\smash{\SetFigFont{7}{8.4}{rm}{\color[rgb]{0,0,0}40}%
}}}
\put(3029,-2999){\makebox(0,0)[b]{\smash{\SetFigFont{7}{8.4}{rm}{\color[rgb]{0,0,0}60}%
}}}
\put(3795,-2999){\makebox(0,0)[b]{\smash{\SetFigFont{7}{8.4}{rm}{\color[rgb]{0,0,0}80}%
}}}
\put(4560,-2999){\makebox(0,0)[b]{\smash{\SetFigFont{7}{8.4}{rm}{\color[rgb]{0,0,0}100}%
}}}
\put(5326,-2999){\makebox(0,0)[b]{\smash{\SetFigFont{7}{8.4}{rm}{\color[rgb]{0,0,0}120}%
}}}
\put(6091,-2999){\makebox(0,0)[b]{\smash{\SetFigFont{7}{8.4}{rm}{\color[rgb]{0,0,0}140}%
}}}
\end{picture}
\vspace*{0.3cm}
\caption{Scaled volume fraction representation of the shadow curve for
Schulz distributions with cutoff $l_m=50$ (solid) and $l_m=100$
(dashed). Notice that, although the discontinuity is much more visible
here than in the density representation
(Fig.~\ref{fig:cloud_shad_50_100_schulz}(b)), the maximum value of
$\rho_1\N$ is not reached at the discontinuity.
}
\label{fig:rho1_50_100_schulz}
\vspace*{-0.5cm}
\efig
The discontinuity in the shadow curves is much more apparent in the
scaled volume fraction representation
(Fig.~\ref{fig:rho1_50_100_schulz}). Notice that now the discontinuity
does not coincide with the point where $\rho_1\N$ reaches its maximum
value, as was the case for the log-normal distribution
(Fig.~\ref{fig:rho1_50_100}).
%
%
%
\begin{figure}[htb]
\begin{center}
\begin{picture}(0,0)%
\includegraphics{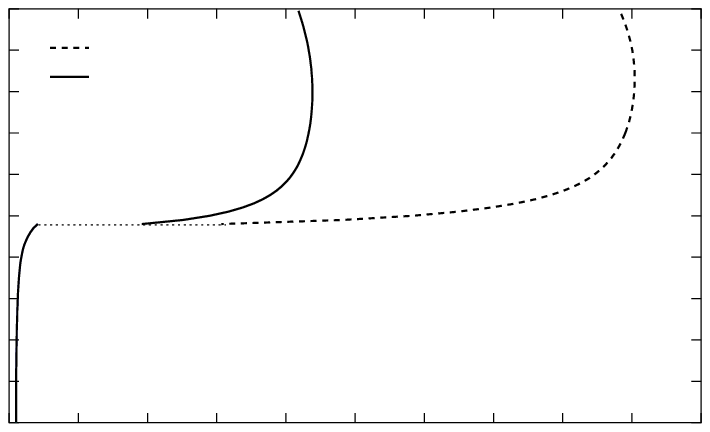}%
\end{picture}%
\setlength{\unitlength}{2447sp}%
\begingroup\makeatletter\ifx\SetFigFont\undefined
\def\x#1#2#3#4#5#6#7\relax{\def\x{#1#2#3#4#5#6}}%
\expandafter\x\fmtname xxxxxx\relax \def\y{splain}%
\ifx\x\y   
\gdef\SetFigFont#1#2#3{%
  \ifnum #1<17\tiny\else \ifnum #1<20\small\else
  \ifnum #1<24\normalsize\else \ifnum #1<29\large\else
  \ifnum #1<34\Large\else \ifnum #1<41\LARGE\else
     \huge\fi\fi\fi\fi\fi\fi
  \csname #3\endcsname}%
\else
\gdef\SetFigFont#1#2#3{\begingroup
  \count@#1\relax \ifnum 25<\count@\count@25\fi
  \def\x{\endgroup\@setsize\SetFigFont{#2pt}}%
  \expandafter\x
    \csname \romannumeral\the\count@ pt\expandafter\endcsname
    \csname @\romannumeral\the\count@ pt\endcsname
  \csname #3\endcsname}%
\fi
\fi\endgroup
\begin{picture}(5952,3874)(151,-3419)
\put(151,-886){\makebox(0,0)[lb]{\smash{\SetFigFont{7}{8.4}{rm}{\color[rgb]{0,0,0}$\sigma$}%
}}}
\put(3412,-3361){\makebox(0,0)[b]{\smash{\SetFigFont{7}{8.4}{rm}{\color[rgb]{0,0,0}$\langle l\rangle$}%
}}}
\put(1501,-196){\makebox(0,0)[lb]{\smash{\SetFigFont{7}{8.4}{rm}{\color[rgb]{0,0,0}$l_m=50$}%
}}}
\put(1501, 29){\makebox(0,0)[lb]{\smash{\SetFigFont{7}{8.4}{rm}{\color[rgb]{0,0,0}$l_m=100$}%
}}}
\put(659,-2875){\makebox(0,0)[rb]{\smash{\SetFigFont{7}{8.4}{rm}{\color[rgb]{0,0,0}0}%
}}}
\put(659,-2555){\makebox(0,0)[rb]{\smash{\SetFigFont{7}{8.4}{rm}{\color[rgb]{0,0,0}0.1}%
}}}
\put(659,-2234){\makebox(0,0)[rb]{\smash{\SetFigFont{7}{8.4}{rm}{\color[rgb]{0,0,0}0.2}%
}}}
\put(659,-1914){\makebox(0,0)[rb]{\smash{\SetFigFont{7}{8.4}{rm}{\color[rgb]{0,0,0}0.3}%
}}}
\put(659,-1593){\makebox(0,0)[rb]{\smash{\SetFigFont{7}{8.4}{rm}{\color[rgb]{0,0,0}0.4}%
}}}
\put(659,-1273){\makebox(0,0)[rb]{\smash{\SetFigFont{7}{8.4}{rm}{\color[rgb]{0,0,0}0.5}%
}}}
\put(659,-953){\makebox(0,0)[rb]{\smash{\SetFigFont{7}{8.4}{rm}{\color[rgb]{0,0,0}0.6}%
}}}
\put(659,-632){\makebox(0,0)[rb]{\smash{\SetFigFont{7}{8.4}{rm}{\color[rgb]{0,0,0}0.7}%
}}}
\put(659,-312){\makebox(0,0)[rb]{\smash{\SetFigFont{7}{8.4}{rm}{\color[rgb]{0,0,0}0.8}%
}}}
\put(659,  9){\makebox(0,0)[rb]{\smash{\SetFigFont{7}{8.4}{rm}{\color[rgb]{0,0,0}0.9}%
}}}
\put(659,329){\makebox(0,0)[rb]{\smash{\SetFigFont{7}{8.4}{rm}{\color[rgb]{0,0,0}1}%
}}}
\put(733,-2999){\makebox(0,0)[b]{\smash{\SetFigFont{7}{8.4}{rm}{\color[rgb]{0,0,0}0}%
}}}
\put(1269,-2999){\makebox(0,0)[b]{\smash{\SetFigFont{7}{8.4}{rm}{\color[rgb]{0,0,0}10}%
}}}
\put(1805,-2999){\makebox(0,0)[b]{\smash{\SetFigFont{7}{8.4}{rm}{\color[rgb]{0,0,0}20}%
}}}
\put(2340,-2999){\makebox(0,0)[b]{\smash{\SetFigFont{7}{8.4}{rm}{\color[rgb]{0,0,0}30}%
}}}
\put(2876,-2999){\makebox(0,0)[b]{\smash{\SetFigFont{7}{8.4}{rm}{\color[rgb]{0,0,0}40}%
}}}
\put(3948,-2999){\makebox(0,0)[b]{\smash{\SetFigFont{7}{8.4}{rm}{\color[rgb]{0,0,0}60}%
}}}
\put(4484,-2999){\makebox(0,0)[b]{\smash{\SetFigFont{7}{8.4}{rm}{\color[rgb]{0,0,0}70}%
}}}
\put(5019,-2999){\makebox(0,0)[b]{\smash{\SetFigFont{7}{8.4}{rm}{\color[rgb]{0,0,0}80}%
}}}
\put(5555,-2999){\makebox(0,0)[b]{\smash{\SetFigFont{7}{8.4}{rm}{\color[rgb]{0,0,0}90}%
}}}
\put(6091,-2999){\makebox(0,0)[b]{\smash{\SetFigFont{7}{8.4}{rm}{\color[rgb]{0,0,0}100}%
}}}
\put(3412,-2999){\makebox(0,0)[b]{\smash{\SetFigFont{7}{8.4}{rm}{\color[rgb]{0,0,0}50}%
}}}
\end{picture}
\vspace*{0.3cm}
\caption{Average rod length in the nematic shadow phase for Schulz
distributions with cutoff $l_m=50$ (solid) and $l_m=100$ (dashed),
plotted against polydispersity $\sigma$ on the $y$-axis.  Above the
discontinuity, the average length jumps to a large value but remains
below $l_m$, implying that the nematic phase
is not yet entirely dominated by the longest rods.
%
}
\label{fig:length_50_100_schulz}
\efig
This is also reflected in a plot of the average rod length in the
nematic shadow against polydispersity
(Fig.~\ref{fig:length_50_100_schulz}), which at the discontinuity in
the shadow curve jumps to a large value but one that is still some way
below $l_m$. In this region the nematic phase thus contains many long
rods, but is not yet entirely dominated by only the longest rods. This
can also be understood by looking back at Fig.~\ref{fig:g0_schulz}:
for values of $z$ just below the discontinuity (corresponding to the
values of $\sigma$ just {\em above} the discontinuity in
Figs.~\ref{fig:rho1_50_100_schulz}
and~\ref{fig:length_50_100_schulz}), $g(0)$ is very close to $z+1$ and
so the overall exponential factor $\exp\{[g(0)-(z+1)]l]\}$ is almost
constant over the range $l=0\ldots l_m$, giving a broad nematic length
distribution~\eqref{eq:fat_condition} dominated by the non-exponential
factors $l^z \angint e^{l[\gt-g(0)]}$.


\subsection{Theory for Schulz distributions with large cutoff}

From the numerical results obtained above, it looks like the case of
the Schulz distribution is actually rather similar to the log-normal
case. In both cases, the presence of the long rods strongly affects
the phase behavior above a certain value of the polydispersity
$\sigma$; the threshold value of $\sigma$ tends to zero as $l_m$
increases for the log-normal case, but appears essentially independent
of $l_m$ for the Schulz distribution. Above the threshold, the nematic
phase is dominated by the long rods present in the system, although
for the Schulz distribution the average length seems to remain
rather below $l_m$. Given these similarities, we now investigate
whether the theory that we developed for the log-normal case can be
extended to the case of the Schulz distribution. As we will see, the
central assumption of dominance of the long rods in the nematic can
still be made self-consistent.
%

If long rods again dominate the density distribution in the nematic
phase, we can repeat all the steps up to Eq.~\eqref{eq:Pi_approx} in
Sec.~\ref{sec:theory_lm}. The same equation for $h\t$ follows, and if
we assume that $\rhoeff=\rho\N l_m^2$ is large we get the
scaling solution $h\t=\hs(\rhoeff\sin\theta)$. Angular integrals can
then again be simplified because only small values of
$\theta\sim1/\rhoeff$ are relevant, leading to
Eq.~\eqref{eq:rhoN_rho}, to the simple result $\Pi\N=3\rho\N$ for the
osmotic pressure in the nematic phase, and to the
expression~\eqref{eq:gnot_asy} for $\gnot=c_1\rho-(8/\pi)\langle
t\rangle_t/l_m$. At this point the different shape of the Schulz
distribution enters: for large $l_m$ we now expect that $\gnot \to
z+1$, rather than $\gnot \to 0$. This then implies that the cloud
point density $\rho$ has a {\em finite} limit for large cutoff, given
by
\beq\label{eq:rho_schulz}
\rho=\frac{z+1}{c_1} = \frac{1}{c_1 \sigma^2}
\eeq
from Eq.~\eqref{eq:sigma}. From the osmotic pressure equality $3\rho\N
= \rho+(c_1/2)\rho^2$ the nematic shadow density then also has a
finite limit,
\beq
\rho\N=\frac{1}{3c_1}\left(\frac{1}{\sigma^2}+\frac{1}{2\sigma^4}\right)
\label{eq:rhoN_schulz}
\eeq
These theoretical predictions for the limiting form of the cloud and
shadow curves in the cutoff-dominated regime are shown by the dotted
lines in Fig.~\ref{fig:cloud_shad_50_100_schulz}, and are
certainly plausible given the numerical results for finite cutoffs.

To obtain the leading terms in the approach of $\rho$ and $\rho\N$ to
their limiting values, and to establish the threshold value of
polydispersity $\sigma$ above which the onset of I-N coexistence is
affected by the presence of the long rods, let us define
$\delta=\gnot-(z+1)$. As pointed out above, Eq.~\eqref{eq:rhoN_rho} still
holds for the Schulz distribution case, but now yields to leading order
\beq\label{eq:z_3}
\rho\N=\frac{c}{l_m^4}\int dl\ l^z e^{\delta l}
\eeq
where we have used $\normparent\l\propto l^z e^{-(z+1)l}$ and
$c$ collects all numerical constants as well as the factor
$\rho/(\rho\N)^2$ which approaches a constant for $l_m\to\infty$ from
Eqs.~\eqref{eq:rho_schulz} and~\eqref{eq:rhoN_schulz}. If $\delta$
converges to zero slowly enough with $l_m\to\infty$ for the product
$\delta l_m$ to diverge, then the exponential factor is dominant in
the integral in Eq.~\eqref{eq:z_3} and we can replace $l^z$ by $l_m^z$
(see App.~\ref{app:schulz_long}) to get
\[
\rho\N=cl_m^{z-4}\frac{e^{\delta l_m}}{\delta}
\]
Rearranging gives, with a new constant $c'$ which contains the
finite limit value of $\rho\N$,
\[
\frac{e^{\delta l_m}}{\delta l_m}=c' l_m^{3-z}
\]
The inverse of the function of $\delta l_m$ on the l.h.s.\ is
asymptotically just a logarithm, yielding
\beq\label{eq:delta_scaling}
\delta\sim (3-z)\frac{\ln l_m}{l_m}
\eeq
For $z<3$, corresponding to $\sigma>1/2$, this result is consistent
with our assumptions: we have $\delta\to0$ and thus $\gnot\to z+1$ as
expected, and also $\delta l_m\to\infty$ as assumed above. From the
convergence of $\delta$ to zero one can then also obtain the approach
of $\rho$ and $\rho\N$ to their limit values: \eg\
$c_1\rho=g(0)+(8/\pi)\langle t\rangle_t/l_m = z+1 + \delta +
\order(1/l_m)$, giving $\rho-(z+1)/c_1 \sim \delta$ to
leading order.

From the above results we can furthermore estimate the average rod
length in the nematic shadow phase for $z<3$ and finite $l_m$. The
exponential factor $\exp(\delta l)$ dominates the nematic density
distribution and so rods in a range of $\order(1/\delta)$ below $l_m$
should contribute to the average length, giving the estimate
\[
\langle l\rangle\N
=l_m\left[1-\order\left(\frac{1}{\delta l_m}\right)\right]
=l_m\left[1-\order\left(\frac{1}{\ln
l_m}\right)\right]
\]
This shows that there are strong logarithmic corrections, consistent
with the fact that in Fig.~\ref{fig:length_50_100_schulz} the average
nematic rod lengths $\langle l\rangle\N$ are still significantly below
$l_m$. By contrast, in the log-normal case, where the role of $\delta$
is played by $g(0)$, the relative corrections to $\langle l\rangle\N$
are $\sim 1/(g(0)l_m) \sim 1/\ln^2 l_m$; even though still
logarithmic, these correction terms are significantly smaller.

Overall, the behavior of ``wide'' Schulz distributions with
$\sigma>1/2$, \ie\ $z<3$, is therefore rather similar to that of
log-normal distributions. We again have $\langle l\rangle\N\to l_m$
for large $l_m$, although now the corrections are rather more
important than in the previous case, and the rods in the nematic are
strongly ordered (since $\rhoeff=\rho\N l_m^2 \sim l_m^2$ diverges for
$l_m\to\infty$). The main difference is the fact that the cloud and
shadow densities, $\rho$ and $\rho\N$, now tend to distinct nonzero
limits for $l_m\to\infty$ rather than to zero: the smaller number of
long rods in the Schulz distribution is not sufficient to induce phase
separation at arbitrarily small densities.

So far we have only covered the regime $z<3$. The case $z=3$ requires a
more careful treatment; here the leading contribution to $\delta$
calculated in Eq.~\eqref{eq:delta_scaling} vanishes, and it turns out
that $\delta$ scales as $1/l_m$, with $\delta l_m$ approaching a
finite limit rather than diverging. Rods in a range $\sim 1/\delta\sim
l_m$ now contribute to the nematic density distribution, which
therefore is no longer dominated by the longest rods alone even for
$l_m\to\infty$; instead, one finds that the distribution approaches a
scaling function of $l/l_m$, with the ratio $\langle l\rangle\N/l_m$
approaching a nontrivial limit value $<1$.

In the case $z>3$, finally, we cannot construct a self-consistent
theory based on the assumption that the nematic phase is dominated by
long rods. Looking at Eq.~\eqref{eq:z_3}, one sees that for $z>3$ {\em
negative} values of $\delta$ would be required to make the r.h.s.\ of
the equation (and thus $\rho\N$) finite for $l_m\to\infty$; with such
negative values of $\delta$, the assumption of dominance of the long
rods in the nematic phase is no longer self-consistent. One might try
the milder assumption that the nematic is dominated by rods which are
long but still short compared to $l_m$, assuming \eg\ $\langle
l\rangle\N\sim l_m^\alpha$ with $\alpha<1$. We found, however, that
this always leads to contradictions. Our conclusion is therefore that
for $z>3$, \ie, $\sigma<1/2$, the nematic phase is dominated by
``short'' rods with lengths not increasing with $l_m$. The
theoretically predicted threshold value of $\sigma$ below which the
cutoff $l_m$ is irrelevant and the presence of long rods in the system
does not significantly affect the phase behavior is therefore
$\sigma=1/2$.

Our numerical results (Fig.~\ref{fig:cloud_shad_50_100_schulz}) show
the kink in the cloud curve, and the corresponding discontinuity in
the shadow curve at $\sigma\approx 0.48$, \ie, close to but slightly
below the theoretically predicted threshold value. Comparing with
Fig.~\ref{fig:g0_schulz}, this corresponds to the fact that, coming
from large $z$, the discontinuous jump in $g(0)$ occurs before the
extrapolation of the solution in the large $z$-regime intersects the
critical line $g(0)=z+1$; consistent with our theory, and
extrapolating by eye, this intersection occurs close to $z=3$. It
therefore appears that for finite $l_m$ the shadow curve jumps to a
cutoff-dependent branch, \ie, a nematic containing many long rods,
already slightly below the asymptotic threshold value $\sigma=1/2$.
The value of $\sigma$ where this jump occurs should then increase
towards $1/2$ as $l_m$ increases.

\section{Conclusion}\label{sec:poly_ons_concl}

We have studied the effect of length polydispersity on the onset of
I-N phase coexistence in the Onsager theory of hard rods. To assess
the possible effects of long rods, two different length distributions
were considered, one with a slowly decaying, ``fat'' tail (log-normal)
and another with an exponentially decaying tail (Schulz).

The log-normal distribution was chosen because it had earlier given
interesting results in the context of homopolymers with chain length
polydispersity~\cite{Solc70,Solc75} and in our previous investigation
of the $\Ptwo$ Onsager model~\cite{SpeSol_p2_fat}, obtained from a
truncation of the angular dependence of the excluded volume of the
Onsager theory. We showed that a length-cutoff $l_m$ needs to be
introduced for fat-tailed distributions such as the log-normal to
avoid divergences in the equations for the onset of phase separation;
the presence of such a cutoff is of course also physically
reasonable. The most striking result from our numerical solution for
the properties of the isotropic cloud and nematic shadow phases is
that the cloud curves show a kink and the shadow curves corresponding
discontinuities: this establishes that for fat-tailed {\em unimodal}
length distributions three-phase I-N-N coexistence occurs. The cloud
and shadow curves show a strong dependence on the cutoff length, with
both moving rapidly to lower densities as the cutoff increases. A plot
of the average rod length in the nematic shadow phase suggested that
it consists almost entirely of the longest rods in the system, \ie,
those of length $l_m$; as a result, the nematic phase also exhibits
very strong orientational ordering.

A theoretical analysis of the limiting behavior for $l_m\to\infty$
confirmed and extended these numerical results. For large cutoffs, the
nematic indeed comprises only the longest rods in the parent length
distribution, and is very strongly ordered. Beyond this, the theory
also predicts that the densities of the isotropic cloud and nematic
shadow phases in fact {\em vanish} (with constant ratio
$\rho/\rho\N=3$) in the limit of infinite cutoff. This rather
surprising result means that even though the {\em average} rod length
in the parent distribution is finite, the fat tail of the distribution
ensures that enough arbitrarily long rods are present to induce phase
separation at any nonzero density. Even though the nematic shadow
density converges to zero for increasing $l_m$, it does so slowly
enough ($\rho\N \sim (\ln^2 l_m)/l_m$) for the rescaled rod {\em
volume fraction} of the nematic to {\em diverge} logarithmically with
$l_m$ ($\rho_1\N \sim \ln^2 l_m$). For any given aspect ratio of rods,
the Onsager second virial approximation thus eventually breaks down as
$l_m$ increases, at the point where the true volume fraction $(D/L_0)
\rho_1\N$ of the nematic shadow becomes non-negligible compared to
unity.

We then studied the case of a Schulz distribution of rod lengths, for
which our previous studies of the simplified
Zwanzig~\cite{ClaCueSeaSolSpe00} and $\Ptwo$ Onsager
models~\cite{SpeSol_p2} showed no I-N-N coexistence and no unusual
behavior in the limit of infinite rod length cutoff. The full Onsager
theory studied here revealed, however, that such effects do indeed
occur: above a threshold value of the polydispersity $\sigma$, the
numerical results show that the nematic density distribution becomes
exponentially divergent for large rod lengths; a finite cutoff $l_m$
again needs to be imposed to get meaningful results. Above the
threshold, the cloud and shadow curves then depend on $l_m$, although
much more weakly than in the log-normal case. The dominance of the
long rods in the nematic shadow phase above the threshold is also
weaker than for the log-normal, with average rod lengths that are
large but significantly below $l_m$. At the threshold itself, a kink
in the cloud curve and a discontinuity in the shadow curve occur,
indicating that even for the ``well-behaved'' Schulz distribution the
Onsager theory predicts a three-phase I-N-N phase coexistence region
in the phase diagram.

We were again able to clarify these results by theoretical analysis of
the limit $l_m\to\infty$. We found that the limiting threshold value
of the polydispersity is $\sigma=1/2$, corresponding to an exponent
$z=3$ in the Schulz distribution, in good agreement with the
numerically calculated thresholds at small cutoffs. Above the
threshold value theory predicts that the average nematic rod length
approaches $l_m$ as in the log-normal case, but now with larger
logarithmic corrections which explain the smaller average lengths
observed numerically. In contrast to the log-normal distribution
the cloud and shadow curves above the threshold approach
{\em finite} limiting values for $l_m\to\infty$. The physical interpretation
of this result is that the smaller number of long rods in the Schulz
distribution is not sufficient to induce phase separation at
arbitrarily small densities.

It is appropriate at this stage to compare the above results with our
earlier analysis of the $\Ptwo$ Onsager
model~\cite{SpeSol_p2_fat}. For the Schulz distribution we have
mentioned already that the $\Ptwo$ Onsager model, in contrast to the
full Onsager theory, predicts no unusual effects (I-N-N coexistence
and cutoff dependences) due to the presence of long rods.  For the
log-normal distribution, the $\Ptwo$ Onsager model does exhibit a
three-phase I-N-N coexistence, with cloud curves showing a kink and
shadow curves a corresponding discontinuity. As in the full Onsager
theory, above the kink/discontinuity the cloud curves are also
strongly dependent on the cutoff value. However, the limiting
behavior of cloud and shadow curves is rather different: both the
densities and the rescaled rod volume fractions of the isotropic cloud
and nematic shadow phases converge to finite, and in fact identical,
limiting values for large cutoff. The nematic was also not dominated
by the longest rods. In fact, the isotropic and nematic phases
differed only through a somewhat larger fraction of long rods
contained in the nematic, and the shorter rods show only negligible
order in the nematic, with the overall orientational order parameter
vanishing in the limit. The enhancement of long rods in the nematic
only becomes detectable in higher order moments of the nematic density
distribution such as $\int dl\ l^n\rho\N\l$, which diverge for
$n>3/2$.

The above differences between the predictions of the $\Ptwo$ Onsager
model and the full Onsager theory can be understood as follows. In the
Onsager theory, the excluded volume of two rods vanishes as the angle
between the rods decreases to zero; this favors strongly ordered
nematics such as the nematic shadow phase dominated by long rods that
we found at the onset of phase coexistence. In the $\Ptwo$ Onsager
model, on the other hand, and indeed in any similar truncation of the
expansion of the kernel $K(\theta,\theta')$ in Legendre
polynomials~\cite{SpeSol_p2}, the excluded volume remains nonzero even
for two rods fully aligned with the nematic axis. This disfavors
nematic phases containing a substantial number of long and strongly
ordered rods. It thus makes sense that the nematic shadow phase even
for the ``fat'' log-normal distribution is predicted to contain only a
small (though enhanced compared to the isotropic phase) fraction of
long rods.

Looking back over our results for the effects of length-polydispersity
in the full Onsager theory, it is clear that all the effects of long
rods that we observe arise from the exponential factor $\exp[lg(0)]$
(see Eq.~\eqref{eq:fat_condition}) which dominates the enhancement of
the nematic shadow phase's density distribution over that of the
isotropic parent phase. Any parent length distribution with a less
than exponentially decaying ``fat'' tail will therefore exhibit
divergences in the nematic distribution, leading to phase behavior
similar to that for the log-normal case. In fact, our theory in
Sec.~\ref{sec:theory_lm} applies to all such fat-tailed
distributions. The Schulz distribution with its exponential tail is
the borderline case, where one cannot predict a priori whether the
presence of long rods will have significant effects. We found that it
does, above a threshold value of the polydispersity. To our knowledge
this is the first time that such an effect has been observed in
polydisperse phase equilibria. In Flory-Huggins theory for
homopolymers with chain length polydispersity, for example, where the
enhancement factor is also a linear exponential (in chain length), no
long rod-effects are found for Schulz
distributions~\cite{Solc75,Solc70}. Finally, for parent rod length
distributions decaying more than exponentially, \eg\ as $\sim
\exp(-l^\alpha)$ with $\alpha>1$, no cutoff dependences are expected
since the nematic density distribution will always be well-behaved for
large lengths. Of course, this does not mean that I-N-N phase
coexistence is excluded for such distributions. Consider for example a
log-normal length distribution modulated by a Gaussian factor
$\exp[-l^2/(2l_m^2)]$ with large $l_m$. As we just saw, there is then
no need for an explicit cutoff. On the other hand, the Gaussian factor
will act as an effective ``soft'' cutoff (hence the notation
$l_m$). For large enough $l_m$ one thus expects phase behavior
qualitatively similar to that discussed above for a ``hard'' cutoff,
including I-N-N phase coexistence signalled by a kink in the cloud
curve.

Above, we have focussed exclusively on the onset of phase coexistence;
both numerically and theoretically the analysis of the phase behavior
inside the coexistence region would be far more challenging. One
question which one would like to answer, for example, concerns the
overall phase diagram topology. On the one hand, the three-phase I-N-N
region could be confined to a narrow density range inside the I-N
coexistence region, as is the case for the $\Ptwo$ Onsager model with
a log-normal length distribution~\cite{SpeSol_p2_fat}.  The
alternative would be for the I-N-N region to extend all the way across
the I-N coexistence region, connecting to the nematic cloud curve and
being bordered by a region of N-N coexistence; this is the behavior
predicted by Onsager theory for bidisperse
systems~\cite{VroLek93,SpeSol_bidisperse_ons}. Apart from a direct
numerical attack on the phase coexistence region, which for now seems
out of reach, clues to an answer could be provided by an approach
based on the Flory lattice model of hard rods~\cite{Flory56_2}. For
the scenarios studied in the past this has yielded results
qualitatively similar to the Onsager
theory~\cite{FloAbe78,AbeFlo78,BirKolPry88}, in spite of the rather
crude treatment of the orientational entropy. Preliminary work shows
that, the limit of thin rods, Flory's excess free energy corresponds
to an excluded volume term which correctly tends to zero for small rod
angles. Together with the full expression for the ideal part of the
free energy this can be shown to give scaling behavior in the limit of
strong ordering (\ie, high density) very similar to that of the full
Onsager theory. This version of the Flory lattice model may therefore
produce predictions that are more in qualitative accord with Onsager
theory than \eg\ the $\Ptwo$ Onsager model. It shares with the latter
the desirable feature of being ``truncatable'', having an excess free energy
that depends only on two moments of the density distribution. This will
allow efficient calculation of phase equilibria using the moment free
energy method~\cite{ClaCueSeaSolSpe00,SpeSol_p2,SolCat98,Warren98,%
SolWarCat01,Sollich02}; work in this direction is in progress.

\appendix

\section{High-density scaling for the monodisperse Onsager theory}
\label{app:high_dens_onsager}

We summarize here the arguments leading to the scaling solution for
the angular distribution in nematic phases at high density, for the
case of monodisperse rods~\cite{VanMul96_high}. The relevant free
energy is obtained from the polydisperse
version~\eqref{eq:ons_free_en} by dropping all $l$-integrations and
setting $l=1$, giving
\bea
f&=&\rho(\ln\rho-1)+\rho\angint P\t\ln
P\t\nonumber\\
&+&\frac{1}{2}\rho^2\angint\thprint\  P\t P(\theta')K(\theta,\theta')\label{eq:mono_ons}
\eea
This expression needs to be minimized with respect to $P\t$, subject
to the normalization condition $\angint P\t=1$ in our usual
notation. One obtains
\beq\label{eq:mono_P}
P\t=\dfrac{e^{\psi(\theta)}}{\int\thprint\ e^{\psi(\theta')}}
\quad 
\psi(\theta)=-\rho\int\thprint\ P\tpr K(\theta,\theta')
\eeq
which is the obvious monodisperse version of
Eq.~\eqref{eq:ons_ptl}. Defining the function
$h\t=\psi(0)-\psi\t$, which obeys $h(0)=0$, this can be
written as
\bea
P\t&=&\dfrac{e^{-h(\theta)}}{\int\thprint\ e^{-h(\theta')}}
\label{eq:mono_ons_P}\\
h(\theta)&=&\rho\int\thprint\ P\tpr \left[K(\theta,\theta')-K(0,\theta')\right]\label{eq:mono_ons_h}
\eea
Now consider the regime of high densities, $\rho\gg 1$. From
Eq.~\eqref{eq:mono_ons_h}, it is clear that for large density, $h\t$
becomes large. $P\t$ then becomes strongly peaked around $\theta=0$ so
that the only non-vanishing contribution to the angular integral in
Eq.~\eqref{eq:mono_ons_h} comes from the range $\theta'\ll 1$. (Here
and in the following we use the symmetry of $P\t$ and $h\t$ under
$\theta\to \pi-\theta$ to restrict all integrations to the range
$\theta=0\ldots \pi/2$.) For $\theta\sim\order(1)\gg \theta'$, we can
then approximate $K(\theta,\theta')\simeq K(\theta,0)=
(8/\pi)\sin\theta$ so that
\beq\label{eq:h_mono_large}
h\t\simeq\frac{8}{\pi}\rho\sin\theta
\eeq
to leading order in $\rho$. This expression will not be valid for
small $\theta$, but suggests that in this regime a scaling solution in
terms of the scaling variable $t=\rho\sin\theta$ could exist, \ie\
$h\t=\hs(t)$. For consistency with Eq.~\eqref{eq:h_mono_large} for
$\theta\sim\order(1)$ (and large $\rho$), the scaling function should then
have the leading asymptotic behavior $\hs(t)=(8/\pi)t$ for large $t$.
Now, written in terms of $\hs(t)$, Eq.~\eqref{eq:mono_ons_h} reads
\beq
\hs(t)=\frac{\int \frac{dt'\ t'}{\sqrt{1-(t'/\rho)^2}} \,
e^{-\hs(t')}\rho\left[K(\theta,\theta')-K(0,\theta')\right]}{\int
\frac{dt'\ t'}{\sqrt{1-(t'/\rho)^2}} \, e^{-\hs(t')}}
\label{eq:h_mono_first}
\eeq
%
where the integrals are over the range $0\ldots \rho$ and
$\theta=\arcsin(t/\rho)$, $\theta'=\arcsin(t'/\rho)$. The key property
that allows one to get a density-independent equation for
$\hs(t)$ is the scaling behavior of the kernel. For finite $t$ and
$t'$ and large $\rho$, one has $\theta\approx t/\rho$, $\theta'\approx
t'/\rho$ and for
such small (and comparable) angles the kernel scales {\em linearly}
with the angles. The product $\rho K(\theta,\theta')$ thus approaches
a finite limit for $\rho\to\infty$,
\bea
& &\rho K(t/\rho,t/\rho')\to \Ks(t,t')=
\sqrt{t^2+t'^2}F\left(\frac{2tt'}{t^2+t'^2}\right),\label{eq:Ks_def}\\
& &F(z)=\frac{8}{\pi}\dint_0^{2\pi}\frac{d\varphi}{2\pi}\ \sqrt{1-z\cos\varphi}\nonumber
\eea
In the same limit we can replace the factors
$[1-(t'/\rho)^2]^{-1/2}$ in Eq.~\eqref{eq:h_mono_first} by 1,
and obtain the scaling equation
\beq\label{eq:scaling_eq_mono}
\hs(t)= \langle \Ks(t,t')-\Ks(0,t') \rangle_{t'}
\eeq
where the average over $t'$ is over the normalized probability
distribution
\beq
\label{eq:Ps}
\Ps(t) = \frac{1}{\gamma} t\,e^{-\hs(t)}, \qquad
\gamma = {\int dt\ t\,e^{-\hs(t)}}
\eeq
of the scaling variable $t$, now running over the range $0\ldots
\infty$ since we have taken $\rho\to\infty$. A plot of the
numerical solution of Eq.~\eqref{eq:scaling_eq_mono} is shown in
Fig.~\ref{fig:h_scaling} and has the expected leading linear behavior
at large $t$.
\bfig
\begin{picture}(0,0)%
\includegraphics{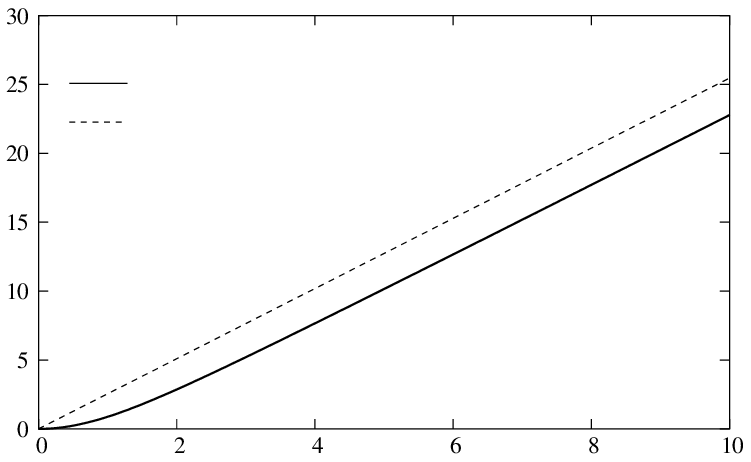}%
\end{picture}%
\setlength{\unitlength}{2447sp}%
\begingroup\makeatletter\ifx\SetFigFont\undefined
\def\x#1#2#3#4#5#6#7\relax{\def\x{#1#2#3#4#5#6}}%
\expandafter\x\fmtname xxxxxx\relax \def\y{splain}%
\ifx\x\y   
\gdef\SetFigFont#1#2#3{%
  \ifnum #1<17\tiny\else \ifnum #1<20\small\else
  \ifnum #1<24\normalsize\else \ifnum #1<29\large\else
  \ifnum #1<34\Large\else \ifnum #1<41\LARGE\else
     \huge\fi\fi\fi\fi\fi\fi
  \csname #3\endcsname}%
\else
\gdef\SetFigFont#1#2#3{\begingroup
  \count@#1\relax \ifnum 25<\count@\count@25\fi
  \def\x{\endgroup\@setsize\SetFigFont{#2pt}}%
  \expandafter\x
    \csname \romannumeral\the\count@ pt\expandafter\endcsname
    \csname @\romannumeral\the\count@ pt\endcsname
  \csname #3\endcsname}%
\fi
\fi\endgroup
\begin{picture}(5757,3713)(1341,-4160)
\put(2401,-1111){\makebox(0,0)[lb]{\smash{\SetFigFont{7}{8.4}{rm}{\color[rgb]{0,0,0}$\hs(t)$}%
}}}
\put(2401,-1411){\makebox(0,0)[lb]{\smash{\SetFigFont{7}{8.4}{rm}{\color[rgb]{0,0,0}$(8/\pi)t$}%
}}}
\put(4201,-4111){\makebox(0,0)[lb]{\smash{\SetFigFont{7}{8.4}{rm}{\color[rgb]{0,0,0}$t$}%
}}}
\end{picture}
\vspace*{0.3cm}
\caption{Numerical solution of Eq.~\eqref{eq:scaling_eq_mono} for the
scaling function $\hs(t)$ (solid) together with
the asymptotic linear behavior at large $t$ (dashed).}
\label{fig:h_scaling}
\efig

In the high-density limit, by the same arguments as above, the
normalization factor for the orientational
distribution~\eqref{eq:mono_ons_P} becomes $\int\thint\,
\exp[-\hs(\rho\sin\theta)] = \gamma/\rho^2$, with $\gamma$ the
normalization factor defined in Eq.~\eqref{eq:Ps}. Thus
%
\beq
P\t=\frac{\rho^2}{\gamma}{e^{-\hs(\rho\sin\theta)}}
\label{eq:P_scaling}
\eeq
Inserting this into the expression~\eqref{eq:mono_ons} for the free
energy and transforming everywhere from $\theta$ to $t$, one has
\bea
f&=&\rho(\ln\rho-1)+2\rho\ln\rho\nonumber\\
&&{}-{}\rho\left[\langle \hs(t)\rangle_t -\ln\gamma\right]
+\frac{1}{2}\rho \langle \Ks(t,t') \rangle_{t,t'}\label{eq:mono_ons_scaling}
\eea
From this it follows that the osmotic pressure for large densities is
simply
\beq\label{eq:Pi_linear}
\Pi=\rho\frac{\partial f}{\partial\rho}-f=3\rho
\eeq
It then also follows that the excess free energy $\fexc$ (the last
term in Eq.~\eqref{eq:mono_ons_scaling}) is just $2\rho$. This can be
seen by comparing the result~\eqref{eq:Pi_linear} with that obtained
via a different route: since $P\t$ is determined to minimize the free
energy, one can evaluate $\Pi=\rho\, \partial\! f/\partial \rho - f$ by
differentiating Eq.~\eqref{eq:mono_ons} while holding $P\t$
constant. Because the excess free energy is quadratic in $\rho$ for
constant $P\t$, this gives $\Pi=\rho+\fexc$ and therefore
$\fexc=2\rho$ by comparison with Eq.~\eqref{eq:Pi_linear}, as claimed.

The result $\fexc=2\rho$ is also useful for deriving an identity which
we use in the main text to show that $\Pi=3\rho$ holds in the nematic
shadow phase at the onset of phase separation in a system with a
fat-tailed length distribution with large cutoff. The excess free
energy is the last term in Eq.~\eqref{eq:mono_ons_scaling}, thus
\beq\label{eq:excl_vol_scaling}
\frac{1}{2}\rho \langle \Ks(t,t') \rangle_{t,t'}=2\rho
\eeq
From Eq.~\eqref{eq:scaling_eq_mono} and $\Ks(0,t')=(8/\pi)t'$
we also have
\bea
\langle\Ks(t,t')\rangle_{t'} &=&
\langle \Ks(t,t')-\Ks(0,t') \rangle_{t'}+\frac{8}{\pi} \langle t'\rangle_{t'}
\nonumber\\
&=&\hs(t)+ \frac{8}{\pi} \langle t \rangle_t
\eea
Inserting into Eq.~\eqref{eq:excl_vol_scaling} we then obtain the desired
identity
\beq\label{eq:mixing_scaling}
\label{eq:scaling_excl_vol_mono}
\langle \hs(t)\rangle_t+\frac{8}{\pi}\langle t \rangle_t=4
\eeq

Notice that arguments very similar to those above apply also to {\em
polydisperse} nematics: one again has a scaling solution $h\t=\hs(t)$
for high density, in terms of the same scaling variable, and this
again leads to $\Pi=3\rho$ and $\fexc=2\rho$. Van Roij and
Mulder~\cite{VanMul96} showed this explicitly for the bidisperse case.

\section{The approximation of dominance of long rods}
\label{app:long_rods_dom}

Our theory in Sec.~\ref{sec:theory_lm} for the onset of phase
separation in systems with fat-tailed length distributions was based
on the assumption that the nematic shadow phase is dominated by the
longest rods in the system. This allowed us to replace terms that
depended weakly on rod length $l$ by their values at the cutoff
$l_m$. We now verify that this assumption is justified in the four
cases where we have used it, namely in
Eqs.~\eqref{eq:rhoN_first},~\eqref{eq:gnot},~\eqref{eq:h_theta}
and~\eqref{eq:Pi_approx}. Let us start from the simplest of these,
Eq.~\eqref{eq:rhoN_first}. Define the angular integral
\[
A(l)=\angint e^{-(l/l_m)h(\theta)}
\]
The density distribution~\eqref{eq:rhoNl_new} in the nematic shadow
phase is then $\rho\N\l=\rho\normparent\l e^{l\gnot}A\l$.  The
normalized nematic length distribution $P\N\l=\rho\N\l/\rho\N$ can be
written as $P\N\l=Q\l A\l/A(l_m)$ if we define
\[
Q\l=\frac{\rho}{\rho\N}\,\normparent\l e^{l\gnot}A(l_m)
\]
Looking back at Eq.~\eqref{eq:rhoN_first}, the assumption of long rod
dominance amounted to replacing the weakly varying factor
$A(l)/A(l_m)$ by $A(l_m)/A(l_m)=1$, effectively substituting $Q(l)$
for $P\N(l)$. To check that this is justified, we need to consider the
unapproximated Eq.~\eqref{eq:rhoNl_new}, which after integration over
$l$ and division by $\rho\N$ reads
\beq\label{eq:rhoN_first_app}
\int dl\ Q\l \frac{A\l}{A(l_m)}=1
\eeq
We effectively approximated this by $\int dl\ Q(l)=1$, so we need to show
that the contribution from the short rods to
Eq.~\eqref{eq:rhoN_first_app} is negligible compared to unity. We will
need the $l$-dependence of $A(l)$ to do this. Restricting the
integration range to $0\ldots \pi/2$ by symmetry, using the scaling
form of $h(\theta)=\hs(t)$ and transforming to the scaling variable
$t=\rhoeff\sin\theta$ gives
\beq
\label{eq:ang_int_short}
A(l)=\rhoeff^{-2}\int_0^{\rhoeff} dt\ \frac{t}{\sqrt{1-(t/\rhoeff)^2}}
e^{-(l/l_m)\hs(t)} 
\eeq
In the limit $l\to 0$ one has $A(l)=1$, and $A(l)$ will remain of this
order while the exponential $\exp[-(l/l_m) \hs(t)]$ is close to unity
even for $t=\rhoeff$. Since $\rhoeff$ is large and $\hs(t)$ linear in
$t$ for large arguments, this gives the criterion $(l/l_m)\rhoeff \sim
1$, or $l\sim l_m/\rhoeff$. Up to this value of $l$, we can
approximate $A(l)\approx 1$. For larger $l$, the factor
$\exp[-(l/l_m)\hs(\rhoeff)]$ is small enough for the integral to be
dominated by values $t\ll \rhoeff$, so that we can set
$[1-(t/\rhoeff)^2]^{1/2}\approx 1$ and extend the upper limit of the
$t$-integral to infinity. The bulk of the integral still comes from
values of $t\gg 1$, however, where $\hs(t)$ is linear, and carrying
out the integral~\eqref{eq:ang_int_short} with this approximation
gives the scaling $A(l) \sim \rhoeff^{-2}(l_m/l)^2$. As $l$ increases,
the range of $t$-values contributing to the integral reduces, and
eventually the quadratic behavior of $\hs(t)$ near $t=0$ leads to
corrections to this scaling. Even at $l=l_m$, however, these effects
are relatively small since $\hs(t)$ is approximately linear even down
to $t\approx 1$ (see Fig.~\ref{fig:h_scaling}); we therefore neglect
them to a first approximation. In summary, we thus have the scaling
$A(l) \sim 1$ for $l\ll l_m/\rhoeff$, and $A(l) \sim
\rhoeff^{-2}(l_m/l)^2$ for $l\gg l_m/\rhoeff$. A sample plot of $A(l)$
evaluated numerically together with our approximation is given in
Fig.~\ref{fig:angint}.
\bfig
\begin{picture}(0,0)%
\includegraphics{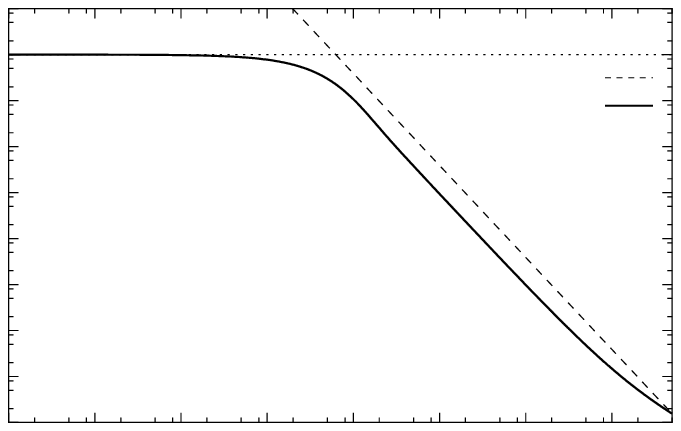}%
\end{picture}%
\setlength{\unitlength}{2447sp}%
\begingroup\makeatletter\ifx\SetFigFont\undefined
\def\x#1#2#3#4#5#6#7\relax{\def\x{#1#2#3#4#5#6}}%
\expandafter\x\fmtname xxxxxx\relax \def\y{splain}%
\ifx\x\y   
\gdef\SetFigFont#1#2#3{%
  \ifnum #1<17\tiny\else \ifnum #1<20\small\else
  \ifnum #1<24\normalsize\else \ifnum #1<29\large\else
  \ifnum #1<34\Large\else \ifnum #1<41\LARGE\else
     \huge\fi\fi\fi\fi\fi\fi
  \csname #3\endcsname}%
\else
\gdef\SetFigFont#1#2#3{\begingroup
  \count@#1\relax \ifnum 25<\count@\count@25\fi
  \def\x{\endgroup\@setsize\SetFigFont{#2pt}}%
  \expandafter\x
    \csname \romannumeral\the\count@ pt\expandafter\endcsname
    \csname @\romannumeral\the\count@ pt\endcsname
  \csname #3\endcsname}%
\fi
\fi\endgroup
\begin{picture}(5231,3790)(881,-3335)
\put(3601,-3286){\makebox(0,0)[lb]{\smash{\SetFigFont{7}{8.4}{rm}{\color[rgb]{0,0,0}$l$}%
}}}
\put(881,-2875){\makebox(0,0)[rb]{\smash{\SetFigFont{7}{8.4}{rm}{\color[rgb]{0,0,0}1e-08}%
}}}
\put(881,-2519){\makebox(0,0)[rb]{\smash{\SetFigFont{7}{8.4}{rm}{\color[rgb]{0,0,0}1e-07}%
}}}
\put(881,-2163){\makebox(0,0)[rb]{\smash{\SetFigFont{7}{8.4}{rm}{\color[rgb]{0,0,0}1e-06}%
}}}
\put(881,-1807){\makebox(0,0)[rb]{\smash{\SetFigFont{7}{8.4}{rm}{\color[rgb]{0,0,0}1e-05}%
}}}
\put(881,-1451){\makebox(0,0)[rb]{\smash{\SetFigFont{7}{8.4}{rm}{\color[rgb]{0,0,0}0.0001}%
}}}
\put(881,-1095){\makebox(0,0)[rb]{\smash{\SetFigFont{7}{8.4}{rm}{\color[rgb]{0,0,0}0.001}%
}}}
\put(881,-739){\makebox(0,0)[rb]{\smash{\SetFigFont{7}{8.4}{rm}{\color[rgb]{0,0,0}0.01}%
}}}
\put(881,-383){\makebox(0,0)[rb]{\smash{\SetFigFont{7}{8.4}{rm}{\color[rgb]{0,0,0}0.1}%
}}}
\put(881,-27){\makebox(0,0)[rb]{\smash{\SetFigFont{7}{8.4}{rm}{\color[rgb]{0,0,0}1}%
}}}
\put(881,329){\makebox(0,0)[rb]{\smash{\SetFigFont{7}{8.4}{rm}{\color[rgb]{0,0,0}10}%
}}}
\put(955,-2999){\makebox(0,0)[b]{\smash{\SetFigFont{7}{8.4}{rm}{\color[rgb]{0,0,0}1e-05}%
}}}
\put(1622,-2999){\makebox(0,0)[b]{\smash{\SetFigFont{7}{8.4}{rm}{\color[rgb]{0,0,0}0.0001}%
}}}
\put(2289,-2999){\makebox(0,0)[b]{\smash{\SetFigFont{7}{8.4}{rm}{\color[rgb]{0,0,0}0.001}%
}}}
\put(2956,-2999){\makebox(0,0)[b]{\smash{\SetFigFont{7}{8.4}{rm}{\color[rgb]{0,0,0}0.01}%
}}}
\put(3623,-2999){\makebox(0,0)[b]{\smash{\SetFigFont{7}{8.4}{rm}{\color[rgb]{0,0,0}0.1}%
}}}
\put(4291,-2999){\makebox(0,0)[b]{\smash{\SetFigFont{7}{8.4}{rm}{\color[rgb]{0,0,0}1}%
}}}
\put(4958,-2999){\makebox(0,0)[b]{\smash{\SetFigFont{7}{8.4}{rm}{\color[rgb]{0,0,0}10}%
}}}
\put(5625,-2999){\makebox(0,0)[b]{\smash{\SetFigFont{7}{8.4}{rm}{\color[rgb]{0,0,0}100}%
}}}
\put(5476,-211){\makebox(0,0)[rb]{\smash{\SetFigFont{7}{8.4}{rm}{\color[rgb]{0,0,0}$\gamma\rhoeff^{-2}(l_m/l)^2$}%
}}}
\put(5476,-436){\makebox(0,0)[rb]{\smash{\SetFigFont{7}{8.4}{rm}{\color[rgb]{0,0,0}$A(l)$}%
}}}
\end{picture}
\vspace*{0.3cm}
\caption{Plot of $A(l)$ against $l$ for $l_m=500$ and
$\rhoeff=6500\, (\approx l_m \ln^2 l_m)$. Notice the
  $\gamma\rhoeff^{-2}(l_m/l)^2$ scaling at intermediate values of $l$ and
  the quasi-constant behavior at small $l$. At $l\sim l_m$
  corrections to the $(l_m/l)^2$ behavior are visible.}
\label{fig:angint}
\vspace*{-0.5cm}
\efig

It follows from the above results that the factor $A(l)/A(l_m)$ in
Eq.~\eqref{eq:rhoN_first_app} is
no larger than $\sim (l_m/l)^2$, even for very small $l$.
The contribution to the $l$-integral from rod lengths $l$ of order
unity is therefore bounded by $\sim \int dl\ l_m^2 l^{-2} Q\l$. For
$l=\order(1)$ 
we can set the factor $\exp[l\gnot]$ in $Q\l$ to one since $\gnot$ is
small (for large $l_m$); the factor $\rho/\rho\N$ is also asymptotically
just an unimportnat constant. The short rods contribution to
Eq.~\eqref{eq:rhoN_first_app} is thus of order $\sim A(l_m) l_m^2 \int
dl\ \normparent\l l^{-2}$ or, if we extend the integral up to
$l=\infty$, $A(l_m) l_m^2 \lav l^{-2}\rav$ with the average taken over
the parent distribution. Now $A(l_m)\sim \rhoeff^{-2}$ (more
precisely, $A(l_m)=\gamma\rhoeff^{-2}$; see before
Eq.~\eqref{eq:P_scaling}) so this contribution scales as
$(l_m/\rhoeff)^2\sim (l_m\rho\N)^{-2}$. For the log-normal length
distribution, $\rho\N \sim (\ln^2 l_m)/l_m$, and so the short rods
contribution $\sim 1/\ln^4 l_m$ to Eq.~\eqref{eq:rhoN_first_app} does
indeed become negligible for large $l_m$, when compared to the long rods
contribution of 1.

Next consider the approximated osmotic pressure
equation~\eqref{eq:Pi_approx}. We need to show that the contribution
of the short rods to the $l$-integral in Eq.~\eqref{eq:Pi_first} is
negligible compared to the long rods part, which we evaluated to be
$2\rho\N$. Dividing by $\rho\N$ to have a quantity to which the long
rods contribute a value of order unity, and discarding factors which
are constant for large $l_m$, we have to consider the integral
\beastar
\int &dl&\
Q\l\frac{l}{l_m}\frac{A\l}{A(l_m)}\biggl\{l_m\left[\gnot-c_1\rho\right]
\\
&-&\left.\frac{1}{A\l}\angint
e^{-(l/l_m)h\t}h\t\right\}
\eeastar
The first term in the curly brackets is easy: the $l$-integral is
proportional to $\int dl\ Q\l(l/l_m)[A\l/A(l_m)]$. Comparing with the
integral $\int dl\ Q\l[A\l/A(l_m)]$ treated above, the short rods
contribution here is suppressed by an additional factor of $l/l_m$ and
so definitely negligible.  The second term, on the other hand, is of
the form
\[
\int dl\ Q\l l \frac{A\l}{A(l_m)}\frac{A'\l}{A\l} =
\int dl\ Q\l l \frac{A\l}{A(l_m)} \frac{d}{dl} \ln A(l)
\]
As we saw above, $A\l$ varies at most as a power-law with $l$, so that
$(d/dl)\ln A(l)\sim 1/l$ and we are again led back to the integral
$\int dl\ Q\l [A\l/A(l_m)]$ for which we showed the dominance of the
long rods above. (This argument applies even in the small-$l$ range 
$l<l_m/\rhoeff$, where $A(l)$ is approximately constant and $(d/dl)\ln
A(l)$ therefore even smaller.)

In Eq.~\eqref{eq:gnot1}, which we approximated by Eq.~\eqref{eq:gnot},
we have a similar $l$-integral that turned out
to scale as $1/l_m$ (see Eq.~\eqref{eq:gnot_asy}). Multiplying then by
$l_m$ to again have a long rods contribution of order unity, and using
$\rho l l_m = (\rho/\rho\N)(l/l_m)\rhoeff \sim (l/l_m)\rhoeff$, we need
to consider the integral
\beq
\int dl\ Q\l \frac{l}{l_m}\frac{A\l}{A(l_m)}\angint
\frac{8}{\pi}\rhoeff\sin\theta\ \frac{e^{-(l/l_m)\hs(\rhoeff\sin\theta)}}{A\l}
\label{eq:aux1}
\eeq
As before, the angular integral will be dominated (with the exception
of very small lengths $l<l_m/\rhoeff$; see below) by the range where
$t=\rhoeff\sin\theta$ is large. In this range $\hs(t)\approx(8/\pi)t$
is linear to a good approximation, so that the angular integral can be
written as 
%
\[
\angint \hs(\rhoeff\sin\theta)
\frac{e^{-(l/l_m)\hs(\rhoeff\sin\theta)}}{A(l)} = \frac{l_m}{A\l}
\]
%
The overall integral~\eqref{eq:aux1} thus becomes $\int dl\ Q\l
l[A\l/A(l_m)](d/dl)\ln A\l$ for which long rods dominance has
been shown already. The contribution from very small $l$-values
$l<l_m/\rhoeff$ needs to be treated separately: here we use the fact
that the angular integral in Eq.~\eqref{eq:aux1} can be viewed as an
average of $(8/\pi)\rhoeff\sin\theta$ over a normalized distribution,
giving a result of at most $\sim \rhoeff$. Using that also $A\l\simeq
1$ in this regime, one needs to integrate $\rhoeff
Q\l(l/l_m)[1/A(l_m)]$ over the range $l=0...l_m/\rhoeff$; the
integrand is bounded there by $Q(l)/A(l_m) \sim \normparent(l)$ and gives a
vanishing integral since the upper limit of the integration range,
$l_m/\rhoeff = 1/(\rho\N l_m) \sim 1/\ln^2 l_m$ vanishes in the limit
$l_m\to\infty$.

Finally, we need to analyse Eq.~\eqref{eq:h_theta1}, which we
approximated by Eq.~\eqref{eq:h_theta}. Let us rewrite
the r.h.s.\ of Eq.~\eqref{eq:h_theta1} as
\[
\frac{\rhoeff}{A(l_m)}
\int\thprint \left[K(\theta,\theta')-K(0,\theta')\right] \int dl\ Q\l
\frac{l}{l_m} e^{-(l/l_m)h(\theta')}
\]
One can view this as an average of the term in square brackets over an
(unnormalized) distribution over $\theta'$. We thus need to show that
the short rods contribution to 
\beq
\int dl\ Q(l) \frac{l}{l_m} e^{-(l/l_m) h(\theta')}
\label{eq:h_correction_int}
\eeq
is negligible compared to the long rods contribution, at least for the
values of $\theta'$ that are in the bulk of this distribution (rather
than the tail). When evaluating the long rods contribution it is not a
priori clear that one can treat the exponential factor
$\exp[-(l/l_m)h(\theta')]$ as weakly varying with $l$. We will see
below that this can nevertheless be justified, so that the long rods
contribution is simply $e^{-h(\theta')}$ as used in
Eq.~\eqref{eq:h_theta}. The short rods contribution to
Eq.~\eqref{eq:h_correction_int} can be estimated by again
approximating $\exp[l\gnot]\approx 1$ and bounding
$\exp[-(l/l_m)h(\theta')]<1$. This yields a contribution of at most
$\int dl\ \normparent\l (l/l_m) A(l_m) < \langle l\rangle A(l_m)/l_m$
with the average again taken over $\normparent\l$ and thus giving
unity.  This is comparable with the long rods term $\exp[-h(\theta')]$
only for values of $\theta'$ such that $e^{-h(\theta')}\sim
A(l_m)/l_m\sim 1/l_m\rhoeff^2\sim \order(l_m^{-3}\ln^{-4} l_m)$. This
means that the corrections due to the short rod integral become
important only where the angular distribution has already decayed to
$\sim 1/l_m^3$ (up to logarithmic terms) of its value at $\theta'=0$,
\ie, far in the tails of the $\theta'$-distribution. We can now also
justify neglecting the $l$-dependence of the term
$\exp[-(l/l_m)h(\theta')]$: at the onset of corrections from the short
rods one has $h(\theta')\sim \ln l_m$. Up to this point, \ie, in the
bulk of the distribution, the ratio $h(\theta')/l_m$ is at most
$\sim(\ln l_m)/l_m$. In the overall exponential factor in
Eq.~\eqref{eq:h_correction_int}, $\exp\{l[\gnot - h(\theta')/l_m]\}$,
this term is therefore still negligible compared to $\gnot\sim(\ln^2
l_m)/l_m$ and the resulting $l$-dependence can be ignored.

\section{The dominance of long rods for the Schulz distribution}
\label{app:schulz_long}

In this appendix we outline briefly how, for the Schulz
distribution~\eqref{eq:schulz_parent}, one can establish that the
assumption of the long rods being dominant in the nematic shadow phase
is again justified (for $z<3$). By way of example we discuss
Eq.~\eqref{eq:rhoN_first}, which is obtained from
Eq.~\eqref{eq:rhoNl_new} if long rods dominate.

As we saw in App.~\ref{app:long_rods_dom}, the scaling of the angular
integral $A\l=\angint e^{-(l/l_m)h(\theta)}$ is
$\rhoeff^{-2}(l_m/l)^2$ for $l>l_m/\rhoeff$, and $A(l)\approx 1$ for
smaller $l$. Inserting this into the nematic density
distribution~\eqref{eq:rhoNl_new}, $\rho\N\l=\rho\normparent\l
e^{l\gnot} A(l)$, using $\rhoeff=\rho\N l_m^2$ and exploiting that
$\rho$ and $\rho\N$ approach constant limits for $l_m\to\infty$, one
has $\rho\N(l) \sim l_m^{-2}l^{z-2}e^{\delta l}$ for $l>l_m/\rhoeff
\sim 1/l_m$, and $\rho\N(l)\sim l^z e^{\delta l}$ for smaller $l$. We
want to show again that the integral of $\rho\N(l)$ over the short
rods (with lengths $l$ of order unity) is negligible compared to the
long rods contribution (which, since $\rho\N = \order(1)$, is of order
unity). In the short rods regime we can approximate $e^{\delta
l}\simeq 1$. The integration of $\rho\N(l)$ over the range $0\ldots
1/l_m$ then gives just $\sim\order(l_m^{-z-1})$ which is negligible
compared to unity for $z\geq 0$. The contribution from the range
$l>1/l_m$ can be bounded by extending the integration not just over
the short rods but in fact up to $l_m$, giving $\sim
l_m^{-2}(l_m^{z-1}-l_m^{1-z})=(l_m^{z-3}-l_m^{-z-1})$ which is again
negligible compared to unity as long as $z<3$. (For $z=1$ the integral
has a logarithmic correction, giving $\sim l_m^{-2}\ln l_m$, but is
still negligible.) As in the log-normal case, the integral over the
nematic density distribution is therefore dominated by the longest
rods, justifying Eq.~\eqref{eq:rhoN_first}.

\bibliographystyle{unsrt}
\bibliography{/home/speranza/Bibliography/references}
\end{document}